Open camera or QR reader and
scan code to access this article
and other resources online.

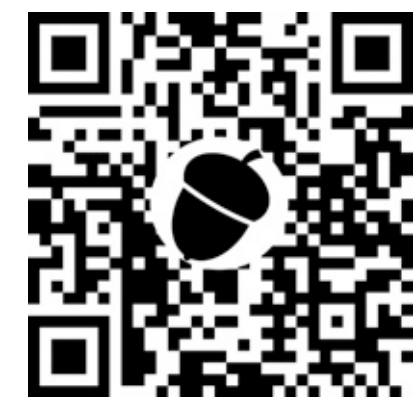

# Simple Lipids Form Stable Higher-Order Structures in Concentrated Sulfuric Acid

Daniel Duzdevich,[1,2,*] Collin Nisler,[1,*] Janusz J. Petkowski,[3,4] William Bains,[5] Caroline K. Kaminsky,[1]
Jack W. Szostak,[1,6] and Sara Seager[7–9]

**Abstract**

Venus has become a target of astrobiological interest because it is physically accessible to direct exploration, unlike exoplanets. So far this interest has been motivated not by the explicit expectation of finding life but rather by a desire to understand the limits of biology. The venusian surface is sterilizing, but the cloud deck includes regions with temperatures and pressures conventionally considered compatible with life. However, the venusian clouds are thought to consist of concentrated sulfuric acid. To determine if any fundamental features of life as we understand them here on Earth could in principle exist in these extreme solvent conditions, we tested several simple lipids for resistance to solvolysis and their ability to form structures in concentrated sulfuric acid. We find that single-chain saturated lipids with sulfate, alcohol, trimethylamine, and phosphonate head groups are resistant to sulfuric acid degradation at room temperature. Furthermore, we find that they form stable higher-order structures typically associated with lipid membranes, micelles, and vesicles. Finally, results from molecular dynamics simulations suggest a molecular explanation for the observed robustness of the lipid structures formed in concentrated sulfuric acid. We conclude with implications for the study of Venus as a target of experimental astrobiology. Key Words: Sulfuric acid—Venus—Lipids—Vesicles. Astrobiology 00, 000–000.

## 1. Introduction

The catalog of confirmed exoplanets continues to grow (Christiansen, 2022), and the James Webb Space Telescope has improved our ability to measure the transmission spectra of exoplanet atmospheres (*e.g.*, Ahrer *et al.*, 2023; Tsai *et al.*, 2023). However, characterizing the physical properties of exoplanets that are relevant to habitability remains extraordinarily challenging. In contrast, our solar system can be explored directly, and Mars, Europa, Enceladus, and more recently Venus are proving of special interest. An astrobiological perspective on the solar system seeks to compare what we know about Earth biology with what we are learning about non-Earth environments through experiments and space missions. Some of our solar system neighbors, like Venus, may appear at first glance to be completely uninhabitable, yet there is nonetheless active speculation about the potential habitability of Venus's sulfuric acid clouds (see, *e.g.*, Bains *et al.*, 2024b, 2021a; Grinspoon and Bullock, 2007; Kotsyurbenko *et al.*, 2024, 2021; Limaye *et al.*, 2018; Mogul *et al.*, 2021a; Morowitz and Sagan, 1967; Patel *et al.*, 2022; Schulze-Makuch and Irwin, 2006; Seager *et al.*, 2021).

[1]Department of Chemistry, Searle Chemistry Laboratory, The University of Chicago, Chicago, Illinois, USA.
[2]Freiburg Institute for Advanced Studies, Albert-Ludwigs-Universität Freiburg, Freiburg im Breisgau, Germany.
[3]Faculty of Environmental Engineering, Wroclaw University of Science and Technology, Wroclaw, Poland.
[4]JJ Scientific, Mazowieckie, Warsaw, Poland.
[5]School of Physics and Astronomy, Cardiff University, Cardiff, United Kingdom.
[6]Howard Hughes Medical Institute, The University of Chicago, Chicago, Illinois, USA.
[7]Department of Earth, Atmospheric and Planetary Sciences, Massachusetts Institute of Technology, Cambridge, Massachusetts, USA.
[8]Department of Physics, Massachusetts Institute of Technology, Cambridge, Massachusetts, USA.
[9]Department of Aeronautics and Astronautics, Massachusetts Institute of Technology, Cambridge, Massachusetts, USA.
*Equal contribution.

However, considering whether a planet is habitable requires an understanding of whether the conditions on that planet are consistent with the conditions required by life (regardless of whether life actually exists there). Earth life remains the basis for such planetary habitability assessment, but the extent to which specifics of Earth life can be considered universal is disputed.

Venus is a critical test case for such astrobiological studies because the clouds consist of liquid concentrated sulfuric acid, which is considered incompatible with life, and contain almost no water, which is considered essential for life (*e.g.*, Hallsworth *et al.*, 2021). Despite these challenges, speculation about life in the clouds of Venus persists because of anomalies in its atmospheric chemistry (*e.g.*, Bains *et al.*, 2021b; Limaye *et al.*, 2018; Mogul *et al.*, 2021a, 2021b; Petkowski *et al.*, 2024b). A rigorous astrobiology research program, centered around organic chemistry, should consider the fundamental features of life as we know it and seek to explore its limits with reference to known environments on other worlds, including Venus.

Organic chemistry in concentrated sulfuric acid is rarely studied yet surprisingly rich, with recent work supporting the notion that complex organic molecules, including amino acids and nucleobases, can be stable in this unusual solvent (Petkowski *et al.*, 2024a, 2025; Seager *et al.*, 2024a, 2023, 2024b; Spacek *et al.*, 2024). These results build on isolated reports from decades ago that considered the chemical behavior of organic molecules in concentrated sulfuric acid long before its importance for planetary science was realized or the sulfuric acid composition of the venusian clouds was known (*e.g.*, Albright *et al.*, 1972; Habeeb, 1961; Miron and Lee, 1963; Reitz *et al.*, 1946; Schumacher and Günther, 1983; Steigman and Shane, 1965; Wagner and von Philipsborn, 1971).

That simple organic molecules can be stable in concentrated sulfuric acid is an interesting observation, and without such stability, life could not be possible in it. But life also requires complex molecular structures for biological function. One fundamental feature of life is cellularity: the differentiation of "inside" (the contents of a cell, including information, molecules, and all their interactions) and "outside" (the environment), in addition to a mechanism for communication and exchange between the two. Cellularity also enables the distinction between encoded genotype (information internal to the cell) and expressed phenotype (the relationship of that information with the environment), a prerequisite for Darwinian evolution. Interpreting cellularity as a feature of life does not necessarily imply Earth biology (a DNA-encoded genotype, for example).

Biology uses lipid membranes to define cells. Unlike proteins and nucleic acids, biological lipids exhibit a broad range of chemistries. They must form effective membranes that can grow, divide in response to the cellular division machinery, and remain stable in the given cell's environment. The primary role of cell membrane lipids is to physically define the cell and thereby demarcate inside from outside. The cell membrane is especially important in extreme environments because it must help maintain the homeostasis of the intracellular environment against otherwise harsh external conditions. Reasoning that with our current understanding of biology, we cannot envision life without cells, we here consider whether simple lipids might be resistant to sulfuric acid, and further, whether they can form higher-order structures such as membranes and vesicles typically considered a prerequisite for life-like phenomena.

The formation of micelles or membranous structures in nonaqueous solvents has attracted little attention. We identified only 11 papers that report the behavior of surfactants and other amphiphilic molecules in concentrated sulfuric acid (McCulloch, 1946; Menger and Jerkunica, 1979; Müller, 1991b, 1991c, 1991a; Müller and Burchard, 1995; Müller and Giersberg, 1992, 1991; Müller and Miethchen, 1988; Steigman and Shane, 1965; Torn and Nathanson, 2002). Several of these studies report the formation of micelles and other lipid structures.

Our research has also been inspired by the extensive literature on the potential role of fatty acids in the emergence of life on Earth (*e.g.*, Apel *et al.*, 2002; Deamer, 2016; Gebicki and Hicks, 1973; Hargreaves and Deamer, 1978; Mansy, 2009). These simple single-chain lipids can form vesicles made of canonical lipid bilayers if the pH of the aqueous solution is near the $pK_a$ of the fatty acid carboxylic acid head groups (typically $\sim$ pH 8 in the context of a membrane) (Kanicky and Shah, 2003). Under these conditions, approximately half of the headgroups will be deprotonated, and the other half protonated, so that charge interaction networks among the headgroups stabilize the hydrophilic bilayer surfaces.

Here we identify several simple lipids that appear to have similar properties in the context of aqueous sulfuric acid solvents in concentrations up to 90% (v/v). We challenge the assumption that concentrated sulfuric acid is inherently incompatible with basic properties of the chemistry of life and specifically show that complex solvent-enclosing lipid vesicles can form, are stable, and can dynamically rearrange and form higher-order structures in concentrated sulfuric acid. While the extreme corrosiveness of sulfuric acid limits our experimental toolkit and ability to comprehensively characterize the biophysical properties of these structures, we posit that they are consistent with lipid bilayer membranes. Our data suggest that micelles, vesicles, oil droplets, and under some conditions extensive membrane networks may be forming. These results signal an opportunity to experimentally push the boundaries of prebiotic chemistry into a regime that is relevant for studying Venus from an astrobiological perspective.

## 2. Materials and Methods

### 2.1. General

Lipids were obtained from Sigma-Aldrich (decyltrimethylamine, octadecyltrimethylammonium, decylphosphonate) or Combi-Blocks (octadecyl sulfate, decylsulfate) at the highest available purity without further purification. Sulfuric acid was from Sigma-Aldrich (nominally 95–99%). Experimental sulfuric acid concentrations are reported as volume/volume (v/v) percentages (volume sulfuric acid/total volume sulfuric acid + water) unless otherwise noted. Water was from a Millipore Milli-Q filtration system with a polisher filtration attachment, rated to dispense water with a resistivity of 18.2 MΩ·cm at 25°C, with total organic carbon ≤5 ppb. Glass vials for sample preparation were from ChemGlass, with PTFE-backed caps. Glassware to be used with sulfuric acid was rinsed with sulfuric acid prior to experimental use. Deuterated NMR reagents and internal standards were from Sigma-Aldrich or Cambridge Isotope Laboratories.

### 2.2. NMR spectroscopy

Samples were prepared by measuring out the appropriate mass of dry lipid in a 4-mL glass vial and adding the previously prepared aqueous sulfuric acid. The vial was sealed with a PTFE-lined cap, and the mixture was gently inverted by hand and allowed to incubate at room temperature for at least 1 h. For extractions, the $CDCl_3$ was added directly to the vial, the mixture inverted gently by hand, and the phases separated over several minutes. The solution was transferred to prerinsed NMR vials with a prerinsed glass Pasteur pipette. NMR spectra were acquired on a Varian INOVA NMR spectrometer operating at 400 MHz for $^{1}H$ spectra and 161 MHz for $^{31}P$ spectra (pulse sequence s2pul). $^{1}H$ spectra in water were referenced to sodium 2,2-dimethyl-2-silapentane-5-sulfonate at 0 ppm. $^{1}H$ spectra in $CDCl_3$ were referenced to tetramethylsilane at 0 ppm. Measurements were locked to $D_2O$ or $CDCl_3$. Chemical shifts ($\delta$) are shown in parts per million (ppm). Spectra were processed and analyzed with MestReNova v14.2.0 (Mestrelab Research) (Willcott, 2009). All full-spectrum NMRs used for preparing figures are in the Appendix to the Supplementary Data.

### 2.3. Dynamic light scattering measurements

Samples were prepared by measuring the appropriate mass of dry lipid in a 4-mL glass vial and adding 2 mL of 70%, 80%, or 90% aqueous sulfuric acid. Samples were placed on a rocker for ~2 h at 20 rpm, after which they were transferred to a glass cuvette with a PTFE cap (Malvern PCS1115), and the day 0 measurements were taken. For the first set of experiments, readings were taken after 24 h and after 5 days. For subsequent analysis of the 50/50 mix of decylsulfate and decyltrimethylamine C10 (SDS, DTMA) or octadecyl sulfate and octadecyltrimethylammonium C18 (SOS, TMO) lipids, measurements were taken hours after preparation, after 24 h, 3 days, and 7 days, unless otherwise indicated. After each measurement, the sample was immediately returned to the rocker. Before each use, vials and cuvettes were cleaned by rinsing with Milli-Q water, detergent, another rinse of Milli-Q water, and then acetone with a vacuum cuvette washer. Measurements were acquired on a Zetasizer Nano ZS at a backscatter detection angle of 175° at 25°C. Three measurements were taken in three runs each for 60 s in the first set of experiments, and for the 50/50 mix of decylsulfate and decyltrimethylamine measurements were taken in three runs for 1000 s per run. The viscosity and refractive index for the indicated sulfuric acid concentrations were taken from the literature (Beyer *et al.*, 1996; Liler, 1971; Rhodes and Barbour, 1923) and used as dispersant parameters in the respective measurements, and attenuator position was determined automatically. Particle sizes were taken from the resultant intensity distribution.

### 2.4. Confocal microscopy

For visualization of 70%, 80%, or 90% aqueous sulfuric acid samples prepared identically as in dynamic light scattering (DLS) experiments, 2 μL of sample was placed in the center of a μ-slide 18-well sterilized, uncoated glass bottom plate (Ibid.). For subsequent additional confocal experiments, vesicles were first prepared in water by measuring the appropriate amount of lipid for a final concentration of 75 mM in a 4-mL glass vial and adding 2 mL of water. After vortexing for ~5 s, 75 mM KOH was added to reproduce conditions used in previous fatty acid studies, and the sample was again vortexed. Aliquots of this sample were added to a μ-slide 18-well glass bottom plate (Ibidi) to five wells, and sulfuric acid was added for a 10%, 50%, 70%, 80%, or 90% final concentration of sulfuric acid. For imaging of samples between glass slides, slides were placed in 1 M NaOH for 5 min, 1 M HCl for 5 min, then washed with deionized water. Two microliters total (lipids plus sulfuric acid) were placed in the center of an imaging spacer (VWR) between slides. For all samples, BODIPY (Sigma) was added in a final concentration of 2.5 μM. Samples were excited at 568 nm and imaged by confocal microscopy on an A1R/Ti setup (Nikon) using a 1.45 NA 100× CFI Plan Apochromat objective.

### 2.5. Molecular dynamics simulations

Membrane patches (100 × 100 Å) of the pure decylsulfate (Sys1), pure decyltrimethylamine (Sys2), and a 50/50 mix of decylsulfate and decyltrimethylamine (Sys3) were generated by the CHARMM-GUI (Jo *et al.*, 2008) membrane builder. TIP3P waters were added 25 Å above and below the bilayer using the autosolvate plugin in visual molecular dynamics (VMD) (Humphrey *et al.*, 1996). In-house Tcl scripts were used to replace all water molecules with sulfuric acid and hydronium ions such that the final concentration of sulfuric acid was ~70% by volume with a net zero charge. Two water molecules were removed for every sulfuric acid added, and one water molecule was removed for each hydronium ion added. The ratio of protonated and deprotonated sulfuric acid molecules was adjusted for each system to give a net zero charge; this resulted in 2672 deprotonated and 1838 protonated sulfuric acid molecules for Sys1, 3228 deprotonated and 1506 protonated for Sys2, and 2894 deprotonated and 1840 protonated for Sys3. The total number of sulfuric acid and hydronium molecules was 4510 and 6012 for Sys1, 4734 and 5788 for Sys2, and 4734 and 5788 for Sys3, respectively. The standard CHARMM36 (Huang and MacKerell, 2013; Vanommeslaeghe and MacKerell, 2012) force field was used for the lipids and sulfuric acid, while parameters for the hydronium ions were taken from previously published results (Sagnella and Voth, 1996). The NAMD 2.14 (Melo *et al.*, 2018; Phillips *et al.*, 2020) program was used to perform all classical and hybrid quantum mechanical/molecular mechanical (QM/MM) simulations, while ORCA (Neese *et al.*, 2020) was used to perform quantum calculations. Analysis was performed, and images were generated using VMD. Results of simulations are reported as significant when two populations exhibited a significant difference when evaluated using a pair sampled *t*-test for means with a hypothesized difference of zero.

To prepare for QM/MM simulations, all systems were minimized for 1000 steps and equilibrated in the *NpT* ensemble at 1 atm using a hybrid Nosé–Hoover Langevin piston method, and temperature was controlled using Langevin dynamics with a damping coefficient of $\gamma = 1$. Nonbonded interactions were switched off at 12 Å, and long-range interactions were calculated using the Particle Mesh Ewald method. Bonds involving hydrogen atoms were fixed using the SHAKE algorithm. At the end of the equilibration, quantum regions were selected to include three interacting lipids: two decylsulfate and one decyltrimethylamine in Sys3, one sulfuric acid molecule and five hydronium molecules for Sys1, two sulfuric acid molecules and one hydronium molecule for Sys2, and three hydronium ions and one sulfuric acid molecule for Sys3. Three such regions were selected for each system, which resulted in nine total QM/MM simulations (Sys1a-c, Sys2a-c, and Sys3a-c). The hydronium ions, the sulfuric acid molecules, and the head groups of all lipids to the C1 carbon atom of the lipid tail were all treated at the QM level (49 atoms total for Sys1a-c, 62 atoms total for Sys2a-c, and 49 atoms total for Sys3a-c), which resulted in a multiplicity of 1 and a net zero charge for all systems. The HF-3c semiempirical Hartree Fock method was used to simulate the QM region (Sure and Grimme, 2013); it was chosen due to its ability to describe large biomolecular systems to an accuracy that approaches large-basis methods at a fraction of the computational cost. Interactions between the QM and MM region were treated with an electrostatic embedding scheme, and covalent bonds split at the QM/MM boundary were treated with a Charge-Shifting method. Charge distribution in the QM region was taken from ORCA and updated at every step. A 0.5 fs timestep was used, and the remaining parameters for the classical region were identical to those used for the equilibration. Self-consistent field electron density and electrostatic potential maps were generated using the orca_plot command and the molecular orbitals output by ORCA and plotted with an isovalue of 0.09.

## 3. Results

### 3.1. Lipid selection

We began by selecting a set of simple lipids that are commercially available at high purity, with saturated hydrophobic tails, and without acid-labile ethers and esters. We limited the tail lengths to 10 or 18. Lipids with C10 and C18 tail lengths were chosen because they should generally be soluble but still hydrophobic enough to promote bilayer formation. Moreover, tail lengths in this range are frequently employed in experimental protocell models, which enables more straightforward comparisons to earlier studies. We sought to include head groups that could in principle form charge–charge interactions, analogous to what is observed with biological lipids, even under heavily protonating conditions. Here, we chose to work with trimethylamine, sulfate, and phosphonate head groups (Caschera *et al.*, 2011;

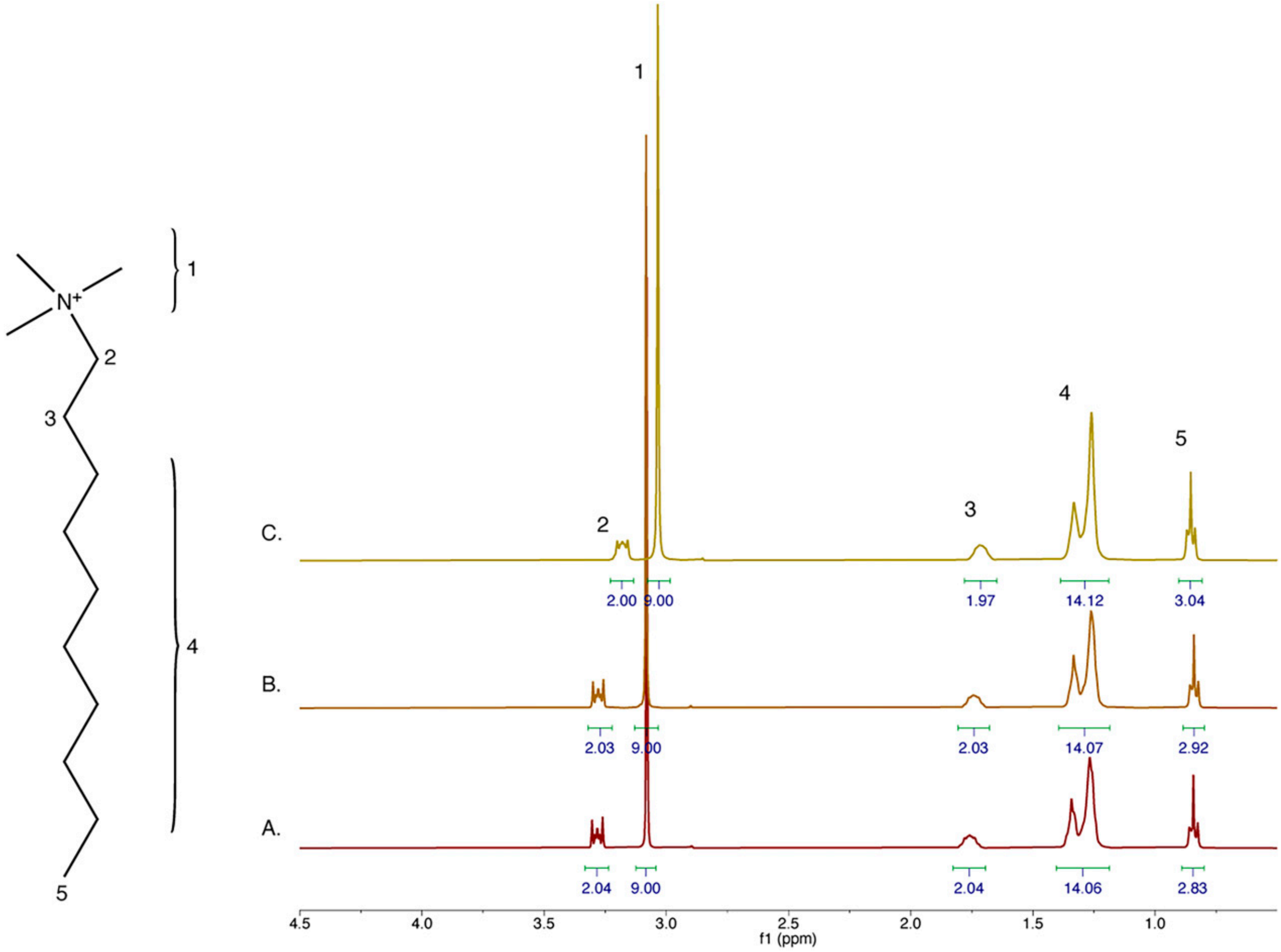

**FIG. 1.** Decyltrimethylamine is resistant to 70% sulfuric acid. $^1$H NMR spectra of 50 mM decyltrimethylamine in (**A**) $D_2O$, (**B**) $D_2O$ + 1% $H_2SO_4$, and (**C**) 30% $D_2O$ + 70% $D_2SO_4$. The measured proton ratios are in agreement with the expected ratios and do not change with increasing sulfuric acid concentration. Note the absence of new peaks. The integrals are normalized to the nine protons assigned to peak 1.

Liu *et al.*, 2018; Walde *et al.*, 1997). We note that other lipids and combinations could be considered in the future.

### 3.2. Nuclear magnetic resonance

We used NMR to test the resistance of each C10 lipid to attack by sulfuric acid at room temperature. At 100 mM, all the selected lipids are poorly soluble in water and concentrated sulfuric acid. However, at 50 mM, decyltrimethylamine ($C_{13}H_{30}N^+$) is soluble in both water and concentrated sulfuric acid (Fig. 1). We incubated 50 mM decyltrimethylamine with $D_2O$, $D_2O$ + 1% $H_2SO_4$, or 30% $D_2O$ + 70% $D_2SO_4$ for at least 1 h before acquiring an $^1H$ spectrum. One percent (v/v) aqueous $H_2SO_4$ corresponds to a measured pH $\cong$ 0.8. We observed no additional peaks or changes to the peaks identified in water under the acidic conditions. Further, the proton ratios remained stable across all conditions. We conclude that decyltrimethylamine is resistant to sulfuric acid in concentrations up to 70%.

Decylsulfate ($C_{10}H_{22}O_4S$) is resistant to 1% $H_2SO_4$ (Fig. 2). However, at 50 mM it is poorly soluble in 70% sulfuric acid and yields low-quality NMR data. Therefore, we incubated decylsulfate in 70% $H_2SO_4$ for 1 h and extracted the mixture into $CDCl_3$ for NMR analysis (Fig. 3). Several new peaks appeared after 70% acid treatment (compare Fig. 2A to Fig. 3A) which we hypothesized arise from decyl alcohol. A spike-in of authentic standard confirmed the assignment (Fig. 3B). We conclude that after 1 h in 70% sulfuric acid, approximately 20% of the sulfate headgroups were hydrolyzed to the alcohol form. This was probably due to acid-catalyzed hydrolysis of protonated sulfate. Decyl alcohol can participate in lipid bilayer formation in the context of other lipids.

Decylphosphonic acid ($C_{10}H_{23}O_3P$) is too poorly soluble in both water and 70% sulfuric acid to generate quality NMR spectra, so we applied the $CDCl_3$ extraction approach for all conditions (Fig. 4). We found that decylphosphonate is stable in up to 70% sulfuric acid, and we note the presence of fully protonated phosphonate at the highest tested acid concentration (Supplementary Fig. S1).

We conclude that sulfuric acid in concentrations up to 70% does not affect decyltrimethylamine, protonates the sulfate head group in decylsulfate, catalyzes the formation of ~20% decyl alcohol, and protonates the phosphonate of decylphosphonate. With the exception of the ~20% decyl alcohol, we observed no acid-catalyzed degradation of any bonds in any of these compounds at room temperature over the 1 h duration of the experiments.

### 3.3. Dynamic light scattering

To determine if phosphonate, trimethylamine, and sulfate lipids form higher-order structures in concentrated sulfuric acid, DLS was employed to obtain the size distribution of lipid aggregates in solution. Initial measurements of 25 mM decylphosphonate, decyltrimethylamine, decylsulfate, an equimolar mix of all three lipids, and a 50/50 equimolar mix of decylphosphonate and decyltrimethylamine were taken in 70% sulfuric acid 24 h after preparing samples and 5 days after preparing samples. Intensity peaks were observed between 100 and 1000 nm for all four samples, which indicates the presence of higher-order structures larger than micelles, regardless of the identity of the lipid head group (Supplementary Fig. S2). (The sizes of micelles and vesicles are determined by the physical properties of the lipids. Simple single-chain lipids form micelles of ~5–20 nm and vesicles of ~100–1000 nm in aqueous solution. The size distributions of micelles and vesicles observed in concentrated sulfuric acid are similar to those observed for fatty acids in water.) In addition, signals were observed above 1000 nm for all samples, which suggests the presence of larger structures. While dust particles or other contaminants could potentially contribute to signals above 1000 nm, microscopy results (see below) showed lipid aggregates at this scale, which suggests that these signals are likely not due to contaminants.

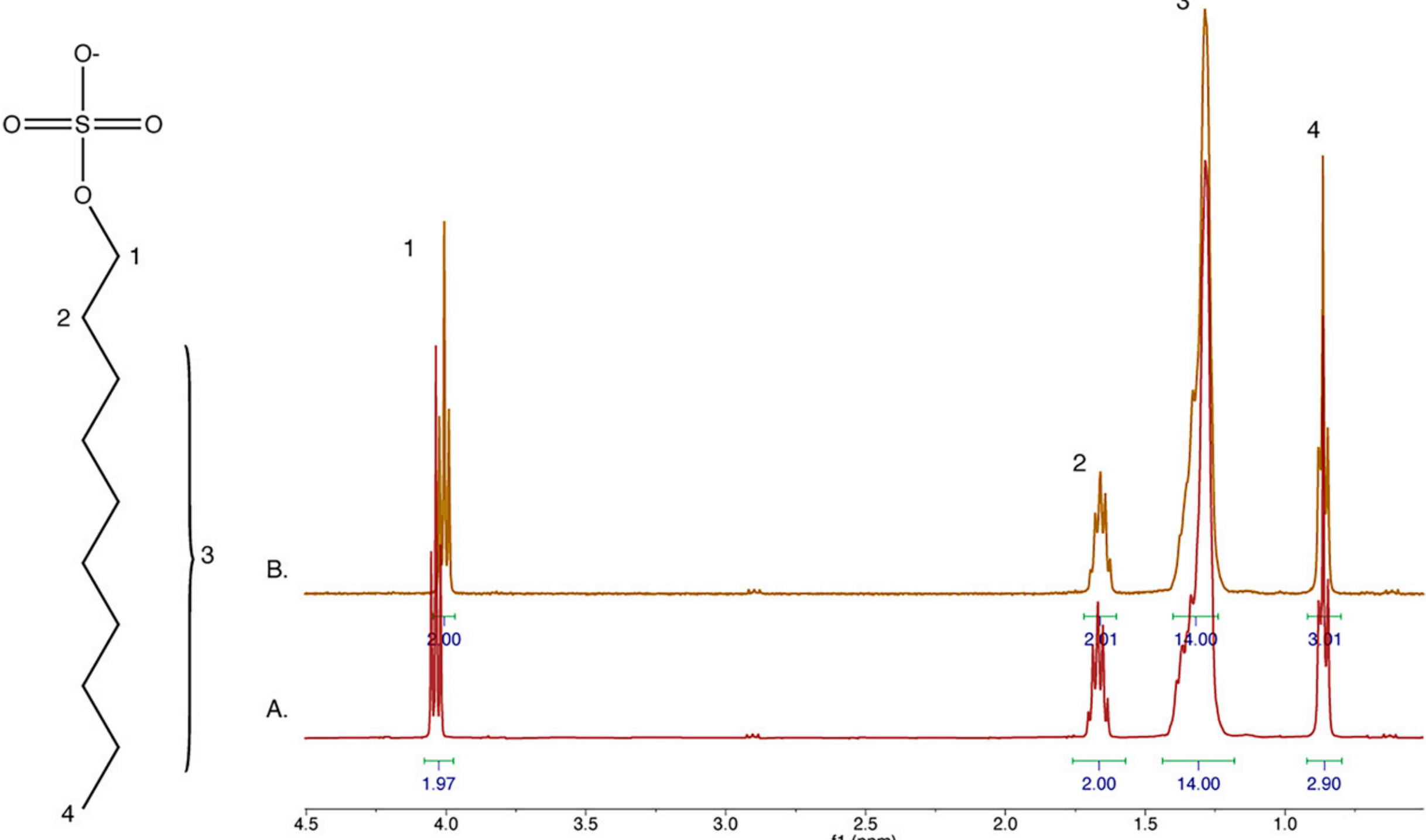

**FIG. 2.** Decylsulfate is resistant to low-concentration sulfuric acid. $^1H$ NMR spectra of 50 mM decylsulfate in (**A**) $D_2O$ and (**B**) $D_2O$ + 1% $H_2SO_4$. The measured proton ratios are in agreement with the expected ratios and do not change with added sulfuric acid. Note the absence of new peaks. The integrals are normalized to the 14 protons assigned to peak 3.

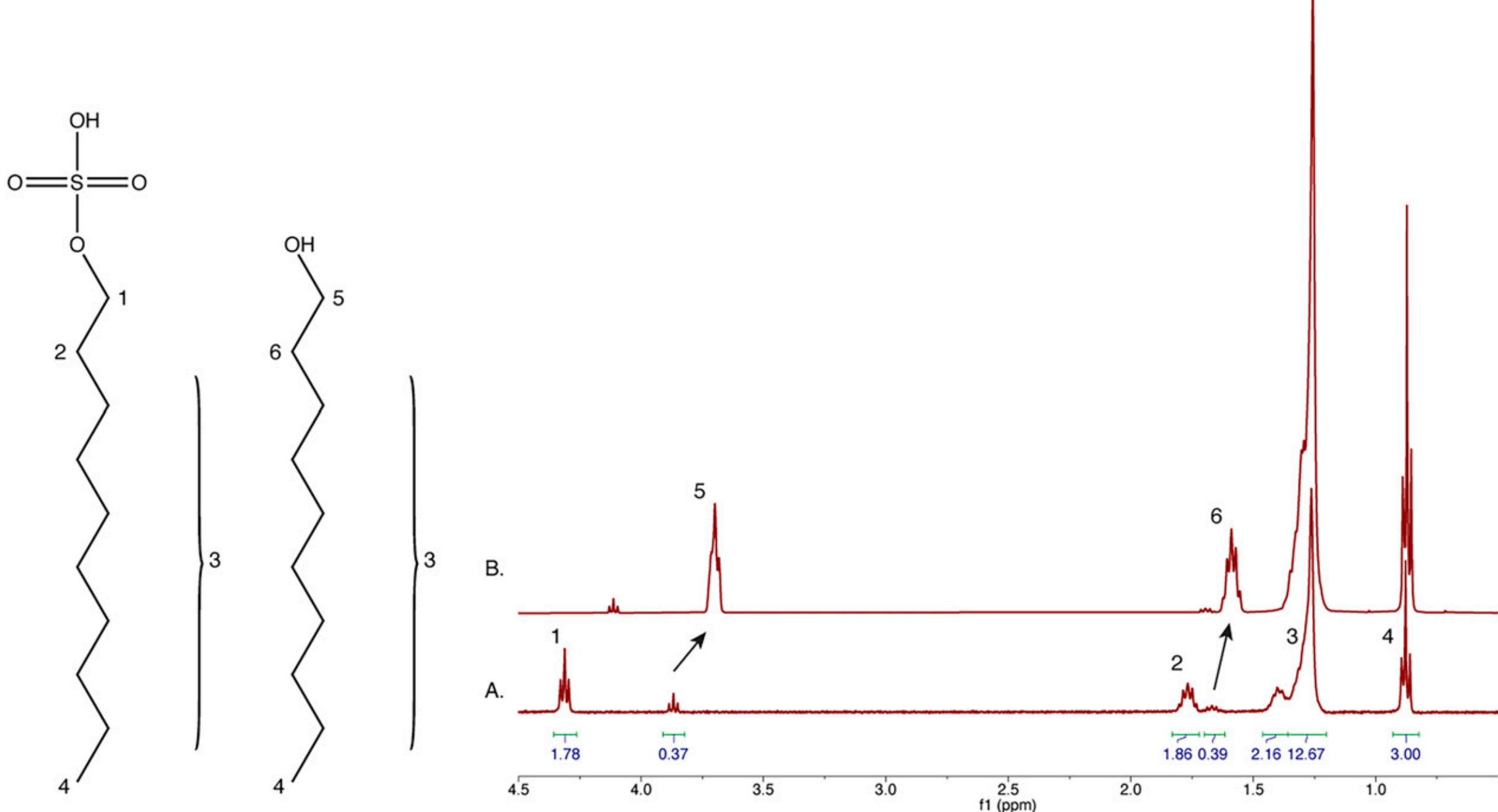

**FIG. 3.** Decylsulfate is largely resistant to 70% sulfuric acid. $^{1}$H NMR spectra of **(A)** 50 mM decylsulfate incubated in 30% $D_2O$ + 70% $H_2SO_4$ for 1 h and extracted into $CDCl_3$, and **(B)** spike-in of decyl alcohol. Compare with Figure 2. We assign the extra peaks upfield of 1 and 2 to the alcohol-adjacent hydrogens in the context of the alcohol, as confirmed by the spike-in of a decyl alcohol standard **(B)**. The peak integral ratios suggest that approximately 20% of the lipids are in the alcohol form upon exposure to concentrated sulfuric acid. The integrals are normalized to the three protons assigned to peak 4.

Decylphosphonate exhibited larger aggregate sizes at both days 2 and 5 compared to all other samples except for the mix of all three lipids at day 2, which suggests that the identity of the headgroup has an effect on the biophysical properties of the lipid structures in sulfuric acid solution. The mix of decylsulfate and decyltrimethylamine exhibited the most stable and least variable signal between measurements, and between days 2 and 5, so lipids with these head groups were selected for further analysis.

To better characterize the changes in size distribution over time and at higher concentrations of sulfuric acid, we prepared a 50/50 equimolar mix of 75 mM C10 SDS-DTMA in 70%, 80%, and 90% sulfuric acid. Furthermore, to determine the effect of chain length on the size distribution of these lipids, a 50/50 equimolar mix of 1 mM C18 SOS-TMO was also tested in 70%, 80%, and 90% sulfuric acid. DLS readings were taken a few hours after preparation (day 0) and again after 1, 3, and 7 days. Three independent replicates were prepared and measured for all samples (Fig. 5; Supplementary Fig. S3). Signals between 100 and 1000 nm, which would roughly correspond to the size of lipid vesicles, were recorded at the peak intensity value. SDS-DTMA (and to a lesser extent SOS-TMO) changed color in concentrated sulfuric acid over the course of 7 days. SDS-DTMA in 90% sulfuric acid became increasingly dark and ultimately almost black (Supplementary Fig. S4). Despite this change in color, the average sizes of lipid structures were not significantly different between chain lengths or over the 7-day incubation as determined by a regression analysis of average size over time ($p > 0.5$). A pair sampled *t*-test for average size between successive measurements and between measurements at different sulfuric acid concentrations showed no significant difference ($p_{\text{correct null}} > 0.05$ for all measurement pairs, with the null hypothesis of 0 difference between means). In addition, with three exceptions (SDS-DTMA in 70% sulfuric acid at days 0 and 3, SOS-TMO in 70% sulfuric acid at day 3; Fig. 5A), the average sizes of SDS-DTMA and SOS-TMO lipid aggregates in 70% or 80% sulfuric acid were not significantly different than the initial 25 mM SDS-DTMA readings (Supplementary Fig. S2E). However, at 90% sulfuric acid, only SDS-DTMA and SOS-TMO at day 0, and SOS-TMO at day 7 exhibited peaks between 100 and 1000 nm on more than one repeat (Supplementary Fig. S5), which suggests that at higher concentrations of sulfuric acid, these structures become more heterogeneous or destabilized. The high viscosity of sulfuric acid may also contribute to the size of vesicle-like structures. Together, the DLS results suggest that the size distribution of lipid structures in concentrated sulfuric acid below 90% is largely unaffected by lipid concentration or chain length and remains stable on the timescale of days.

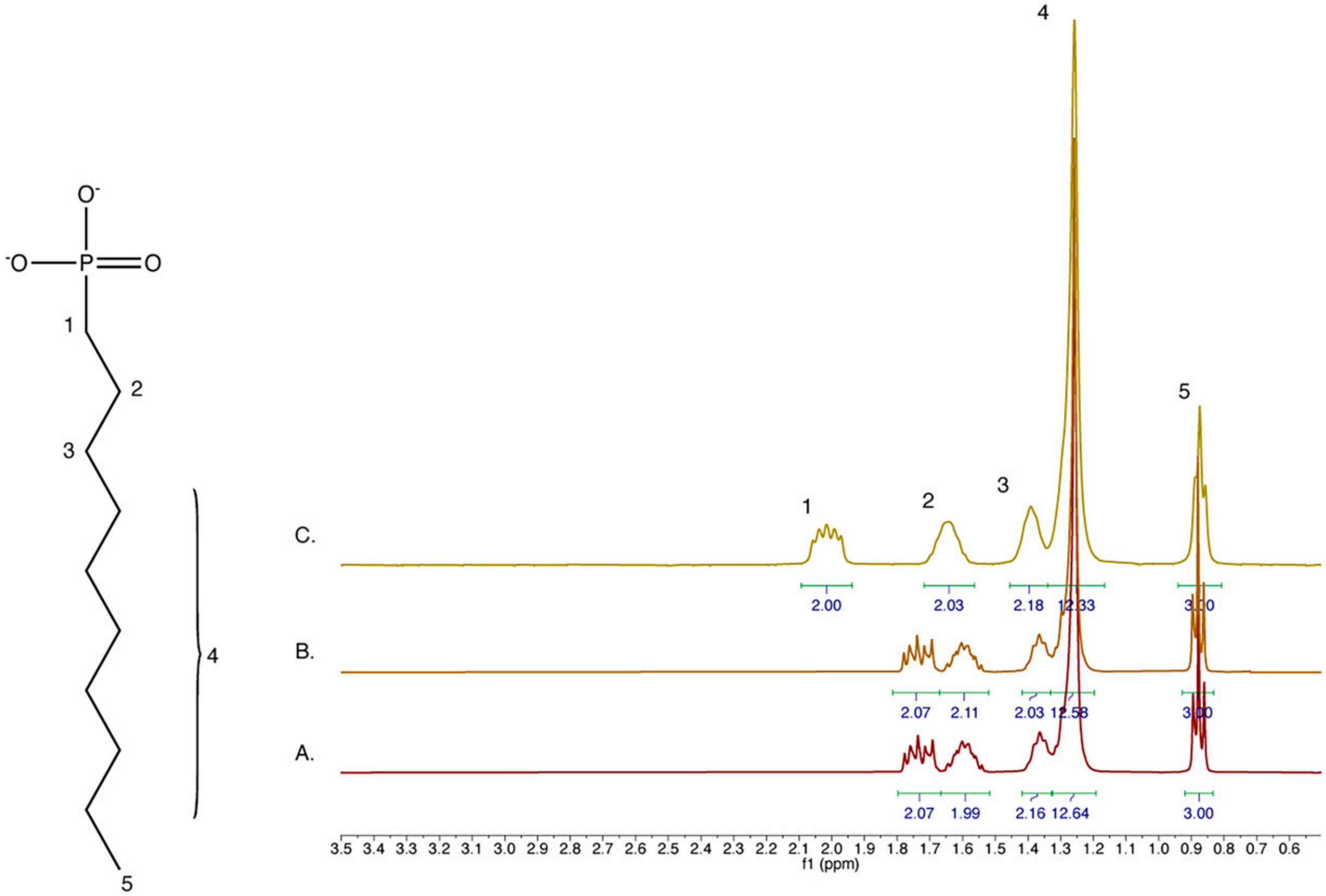

**FIG. 4.** Decylphosphonate is resistant to 70% sulfuric acid. $^1$H NMR spectra of 50 mM decylphosphonate incubated in **(A)** $D_2O$, **(B)** $D_2O$ + 1% $H_2SO_4$, and **(C)** 30% $D_2O$ + 70% $H_2SO_4$ for 1 h and extracted into $CDCl_3$. The measured proton ratios are in agreement with the expected ratios and do not change with increasing sulfuric acid concentration. Note the absence of new peaks. Peaks 1 and 2 shift downfield as the phosphonate is protonated (see Supplementary Fig. S1). The integrals are normalized to the three protons assigned to peak 5.

### 3.4. Confocal microscopy

Our DLS measurements suggest that structured lipid aggregates can form in concentrated sulfuric acid. To determine the morphology of these structures, we used confocal microscopy to visualize 4,4-difluoro-4-bora-3a,4a-diaza-*s*-indacene (BODIPY) stained lipids. Samples were prepared identically as in the DLS experiments and placed in the center of an 18-well uncoated, sterilized glass-bottom slide such that samples did not contact the walls of the chamber. We then added 2.5 μM BODIPY to each sample. Surprisingly, the organic BODIPY dye showed robust fluorescence despite the concentrated sulfuric acid solvent. In all samples, vesicle-like structures of various sizes and lipid morphology were observed (Fig. 6), which corresponded to the range of sizes observed in the DLS signals. The presence of these structures is strictly dependent on the addition of lipids; thus, this eliminates the possibility that the observations were an artifact of sulfuric acid interaction with the slide surface. In 70% and 80% sulfuric acid, vesicle-like structures were distributed throughout the sample, while in 90% sulfuric acid, they were segregated to the periphery of the sample. This suggests a possible gradient in sulfuric acid concentration within the sample with vesicles only able to form in a local region of lower acid concentration. This may have been an artifact of the hygroscopic sulfuric acid absorbing moisture from the air. These results suggest that SDS-DTMA lipids can form membranous, vesicle-like structures at sulfuric acid concentrations up to 90%. Additionally, all samples exhibited aggregates of widely varying sizes up to several microns, which agrees with the polydisperse samples suggested by the DLS readings (Fig. 5; Supplementary Fig. S2).

Next, we tested the ability of preformed SDS-DTMA vesicles in water to resist destabilization upon exposure to sulfuric acid. Lipids were prepared in water, and aliquots from this initial sample were added to individual glass-bottom wells. Sulfuric acid was added to each well for a final concentration of 10%, 50%, 70%, 80%, or 90%, and BODIPY was added to a final concentration of 2.5 μM. Images were taken shortly after preparing samples and 3 days after prepaing samples (Fig. 7). In the sample without sulfuric acid, multilamellar structures were abundant (Fig. 7A) and adopted a relatively consistent size, which indicates that the SDS-DTMA mix readily forms vesicle-like structures in an aqueous environment. In the presence of 10% or 50% sulfuric acid, a change in morphology was observed relative to the 0% (water-only) sample (Fig. 7B and C). These samples exhibited a more heterogeneous composition, with the presence of generally thicker membranes and a more irregular spatial distribution of aggregates and nonmembranous droplets. Vesicle-like structures were present in both 10% and 50% sulfuric acid at day 0 and day 3. A further change in morphology was observed in 70% and 80% sulfuric acid (Fig. 7D and E). Mostly aggregates or droplets were observed at day 0, but between day 0 and day 3, these droplets coalesced to form a series of multilamellar structures, which resemble vesicles, with an apparent internal volume. A three-dimensional image of these structures shows that they adopt a distinct ovoid shape (Fig. 7E, final panel). Finally, at 90% sulfuric acid, no membranous structures were observed at day 0 or 3 (Fig. 7F). These results suggest that vesicles formed by SDS-DTMA lipids remain stable when exposed to the addition of sulfuric acid below 90%.

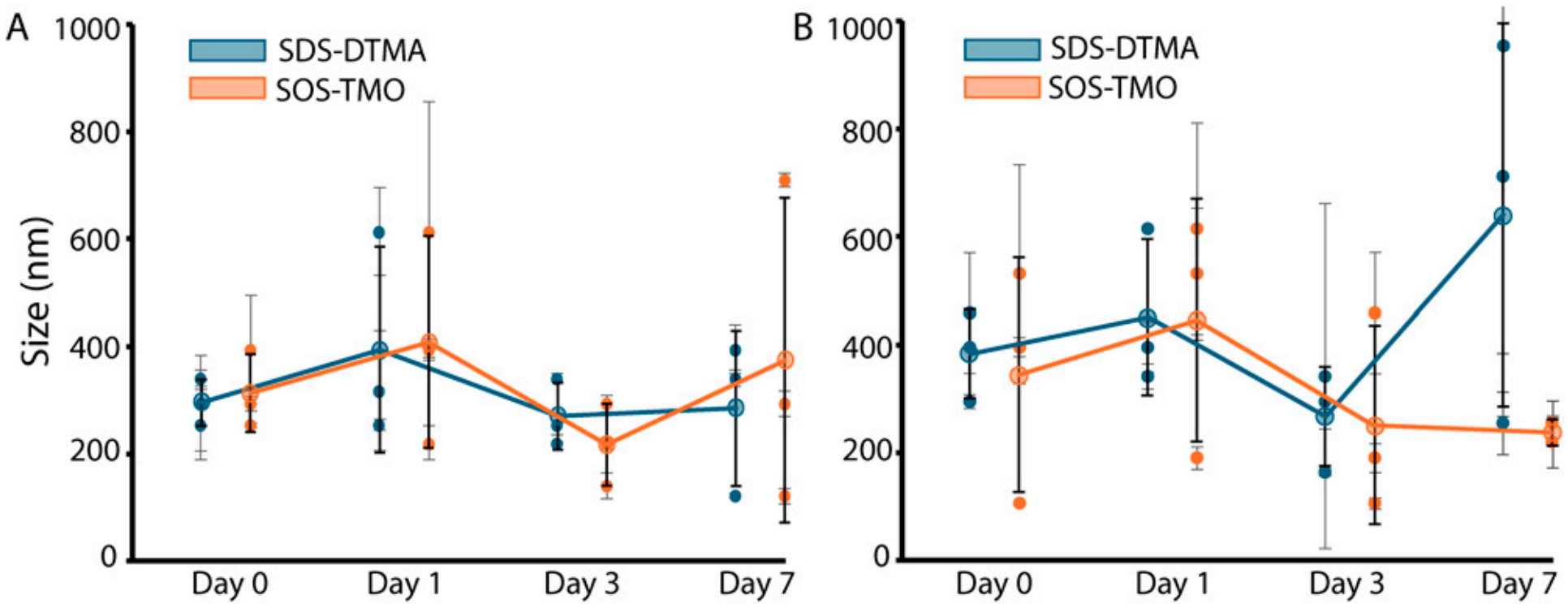

**FIG. 5.** Lipids form stable structures in concentrated sulfuric acid. DLS readings of 75 mM decylsulfate and decyltrimethylamine (SDS-DTMA, blue) and 1 mM octadecyl sulfate and octadecyltrimethylammonium (SOS-TMO, orange) lipids prepared in **(A)** 70% sulfuric acid and **(B)** 80% sulfuric acid. Average values of independent replicates for each lipid preparation are shown by large transparent circles, individual values are indicated by small solid circles, and standard deviations are shown in black calculated from independent replicates. Standard deviations for individual data points, obtained through a standard fit to the corresponding size distribution, are shown in gray lines. These results show that SDS-DTMA and SOS-TMO lipids form structures in 70% and 80% sulfuric acid on the same scale as fatty acid vesicles, and that no significant change in size is observed over the course of a week. DLS, dynamic light scattering.

In all samples besides those in 0% or 90% sulfuric acid, a pronounced change in membrane morphology was observed between days 0 and 3 (Fig. 7). This change could be the result of a structural rearrangement of lipids within the membrane due to the change in solvent properties, the heat released upon the addition of sulfuric acid, or osmotic shock upon addition of sulfuric acid. To determine if heat can induce a morphology change, 75 mM SDS-DTMA vesicles prepared as above were subjected to a heat shock of 65°C for 1 h. While some change may have occurred (Supplementary Fig. S6A), the size and morphology appeared similar to that without heat shock (Fig. 7A). Additionally, adding water in the same volume and ratio to samples prepared in 70%, 80%, and 90% sulfuric acid did not result in the morphological change observed upon addition of the same volume of sulfuric acid (Fig. 7). These results indicate that the changes observed between days 0 and 3 in the presence of sulfuric acid were likely due to changes in lipid–solvent interactions, not the heat released upon addition of sulfuric acid or osmotic shock.

To ensure the structures observed in Figure 7 were not the result of interaction between sulfuric acid and the walls of the sample well, and to better characterize the temporal determinant of the observed morphology changes, we prepared SDS-DTMA samples in 70%, 80%, and 90% sulfuric acid as above but placed between glass slides such that sulfuric acid only contacted clean glass. Images were taken on days 0, 1, 3, and 7 (Fig. 8A–D). All samples exhibited membranous vesicle-like structures of various sizes and morphologies with internal volumes (Fig. 8E and F). As in Figure 7, a change in morphology was observed after day 0 for all three samples. For all three samples with sulfuric acid, but not in the aqueous sample, a large, dense, and three-dimensional cluster of vesicle-like structures with diverse sizes and lamellarity was observed on day 3 or 7 (Fig. 8B–D, final frame). This clustering of vesicles may indicate a local exclusion of sulfuric acid in the interior of the cluster that has the effect of protecting the vesicles from disruption to their membranes. The mechanism behind the clustering could be positive surface charges bridged by sulfate anions, but further analysis will be required to explore this possibility. Vesicles were present outside of the large cluster structures in all samples, both adhered to the surface and free-floating, which indicates that while clustering among vesicles may provide some degree of protection, vesicles can exist individually in concentrated sulfuric acid. Additionally, elongation of the membrane up to tens of microns in length was seen in all three sulfuric acid samples but not in aqueous solvent (Fig. 8G), often with associated pearling along the elongated segment. This behavior has been observed upon growth and mechanical perturbation of oleic acid vesicles in aqueous solvent and was proposed as a mechanism for vesicle growth and division (Zhu and Szostak, 2009). These results suggest that a similar cycle can be achieved in concentrated sulfuric acid with SDS-DTMA vesicles.

### *3.5. QM/MM modeling of the lipid bilayer membrane*

Results of the confocal imaging suggest that SDS-DTMA membranes remain stable in the presence of sulfuric acid. To determine how the head groups of these lipids interact with each other and with concentrated sulfuric acid to maintain a stable membrane, three independent QM/MM simulations were run for a total of 125 ps per system on a membrane patch composed of DTMA, SDS, or a 50/50 mix of SDS-DTMA lipids for a total of nine simulations (Supplementary Fig. S7). For each system, three lipid head groups and interacting solvent molecules were treated at the QM level. At the end of each simulation, electron density and charge distribution were calculated. As expected, SDS head groups exhibited strong negative charge and DTMA head groups strong positive charge (Fig. 9A–C); this results in a repulsion between head groups and an attraction between oppositely charged head groups. The attractive forces result in a more energetically stable system with SDS-DTMA. Because the bilayer surface is perpendicular to the *z*-axis, deviation of the plane formed by the three lipid heads from the *x*–*y* plane can indicate the propensity for membrane deformations to form, which makes the membrane more susceptible to solvent penetration. The angle between an axis perpendicular to the plane formed by the three lipid heads and the *z*-axis was tracked during all simulations, and the SDS-DTMA and DTMA systems exhibited a significantly lower angle than SDS-only (Fig. 9D, left panel; $p < 0.05$). This indicates that the mixed lipid and pure DTMA systems remained most planar.

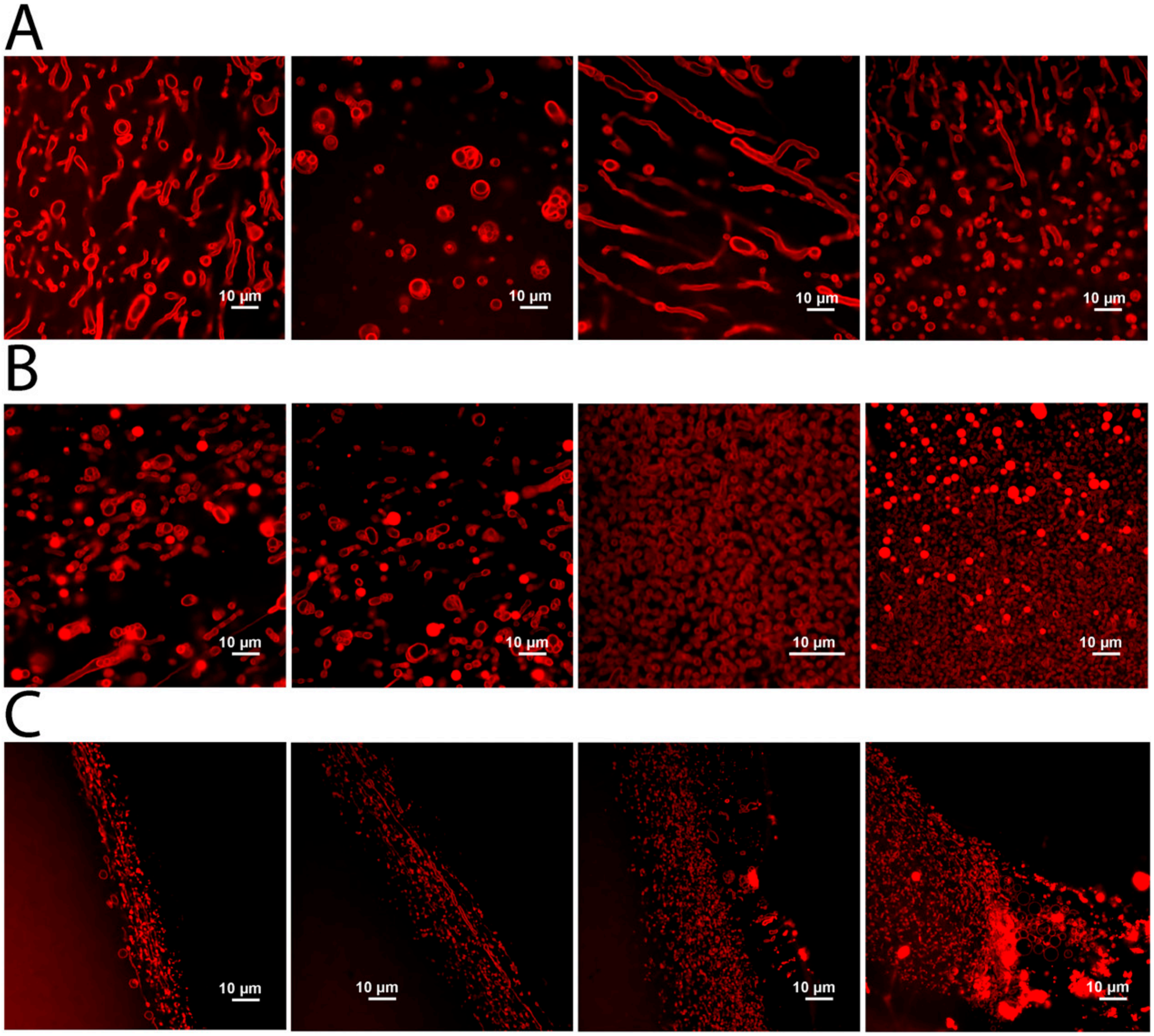

**FIG. 6.** Vesicle-like structures form after the addition of concentrated sulfuric acid to solid lipids. Confocal images of BODIPY-stained SDS-DTMA lipids from samples prepared in **(A)** 70%, **(B)** 80%, and **(C)** 90% sulfuric acid. Vesicle-like structures of different sizes and other lipid morphologies are observed in all samples. Each panel is a different region of the same sample, taken on the same day. These results indicate that the signals observed in the DLS readings are due to the presence of lipid-based structures. SDS-DTMA, decylsulfate and decyltrimethylamine.

Next, average distance between the center of mass of head groups was monitored during simulations. The SDS-DTMA system exhibited a significantly lower average distance between head groups (Fig. 9D, right panel; $p < 0.05$). These results suggest that, based on the dynamics of the lipid groups and their interactions, the stronger interactions in the mixed SDS-DTMA system make the bilayer more resistant to disruption by sulfuric acid. This disruption likely comes from perforation by sulfuric acid and hydronium ions, which compromises membrane continuity. In addition to interlipid interactions, interactions between lipids and solvent also suggest a mechanism by which a mix of DTMA and SDS lipids are more robust to destabilization by sulfuric acid. In the SDS-only systems, interactions between lipids were not direct but facilitated by hydronium ions (Fig. 9A; Fig. 9E, left). This in turn leads to the penetration of hydronium molecules into the bilayer, which was not observed in the DTMA-only or SDS-DTMA systems (Fig. 9B and C). Furthermore, while the total number of hydrogen bonds was highest in the SDS-only system (Fig. 9E and F, left panel), almost none of the hydrogen bonds in the SDS-only system came from interaction between head groups throughout the simulations (Fig. 9E and F, right panel). In the SDS-DTMA system, there is a robust network of hydrogen bonds between both the solvent and head groups and between head groups themselves, preventing solvent molecules from penetrating beyond the head groups (Fig. 9E). Very few hydrogen bonds, either interlipid or lipid–solvent, were observed in the DTMA-only system (Fig. 9F). Collectively, the simulation results suggest the SDS-DTMA system forms stronger interlipid interactions and a more extensive hydrogen bonding network between lipids and solvent facilitated by a more even charge distribution across lipids, which results in a more stable bilayer.

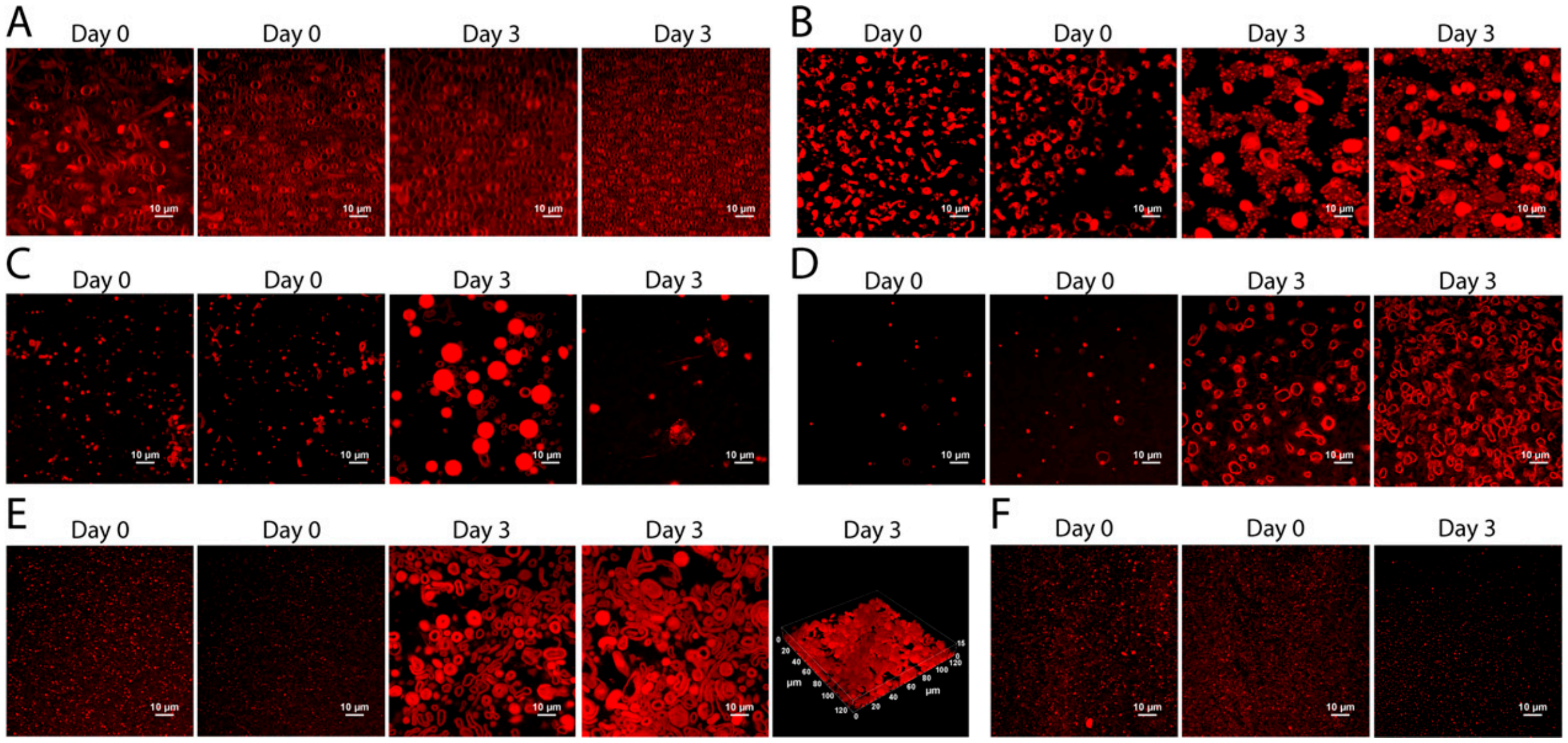

**FIG. 7.** Vesicle-like structures formed in water persist after the addition of sulfuric acid. Confocal images of BODIPY-stained SDS-DTMA lipids at day 0 and day 3 at **(A)** 0%, **(B)** 10%, **(C)** 50%, **(D)** 70%, **(E)** 80%, or **(F)** 90% sulfuric acid. The final frame of **(E)** is a three-dimensional z-stack image of that shown in the fourth panel. These results show that preformed vesicle-like structures persist after the addition of concentrated sulfuric acid.

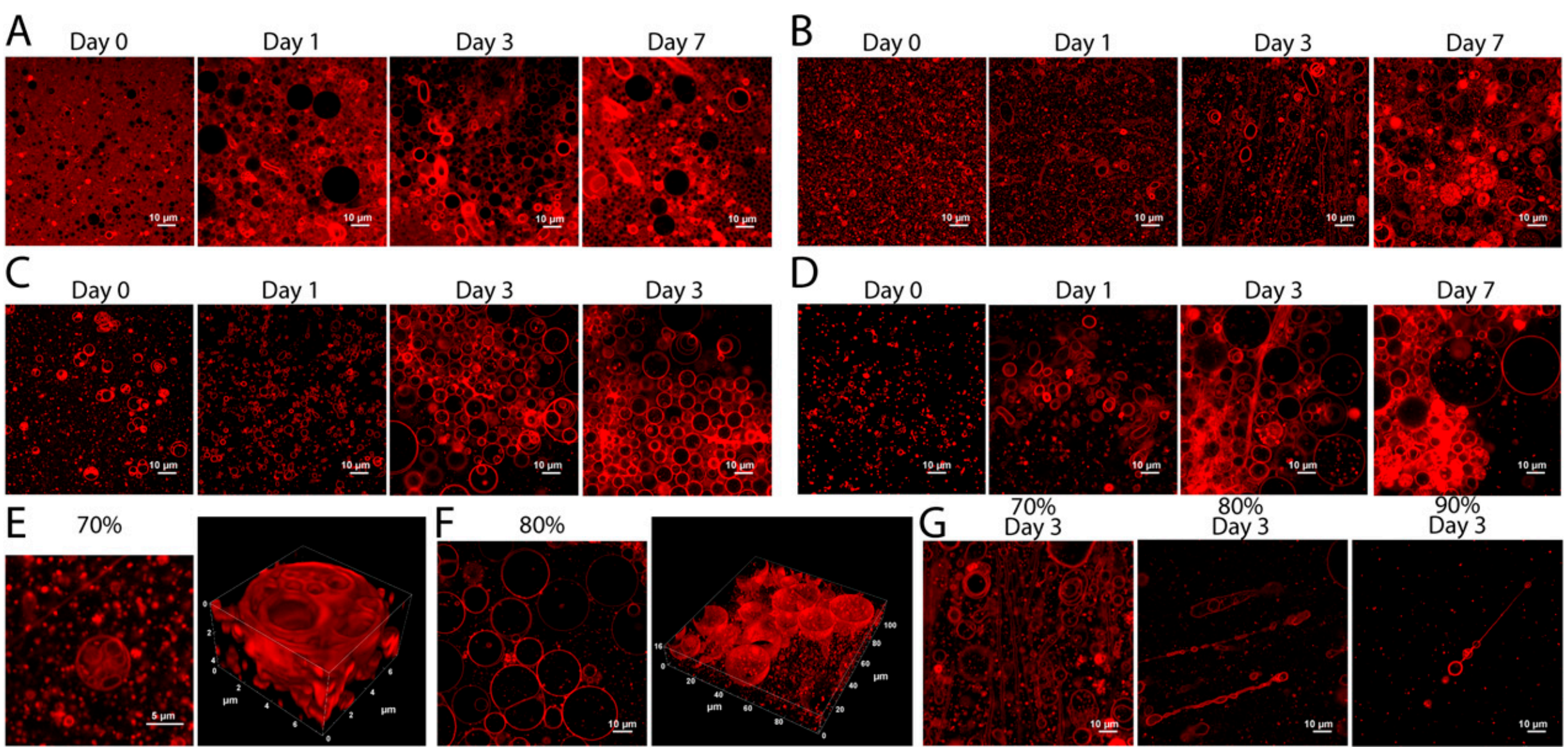

**FIG. 8.** Vesicle-like structures persist in sulfuric acid over the course of 7 days. Confocal images of BODIPY-stained SDS-DTMA lipids at day 0, day 1, day 3, and day 7 at **(A)** 0%, **(B)** 70%, **(C)** 80%, or **(D)** 90% sulfuric acid. Z-stack images showing the structure and internal volume of vesicle-like structures in **(E)** 70% or **(F)** 80% sulfuric acid. **(G)** Pearling and elongation of membranes were widespread in all samples and are shown for 70%, 80%, and 90% sulfuric acid. These images show that morphologically diverse vesicle-like structures are stable in sulfuric acid over the course of 7 days.

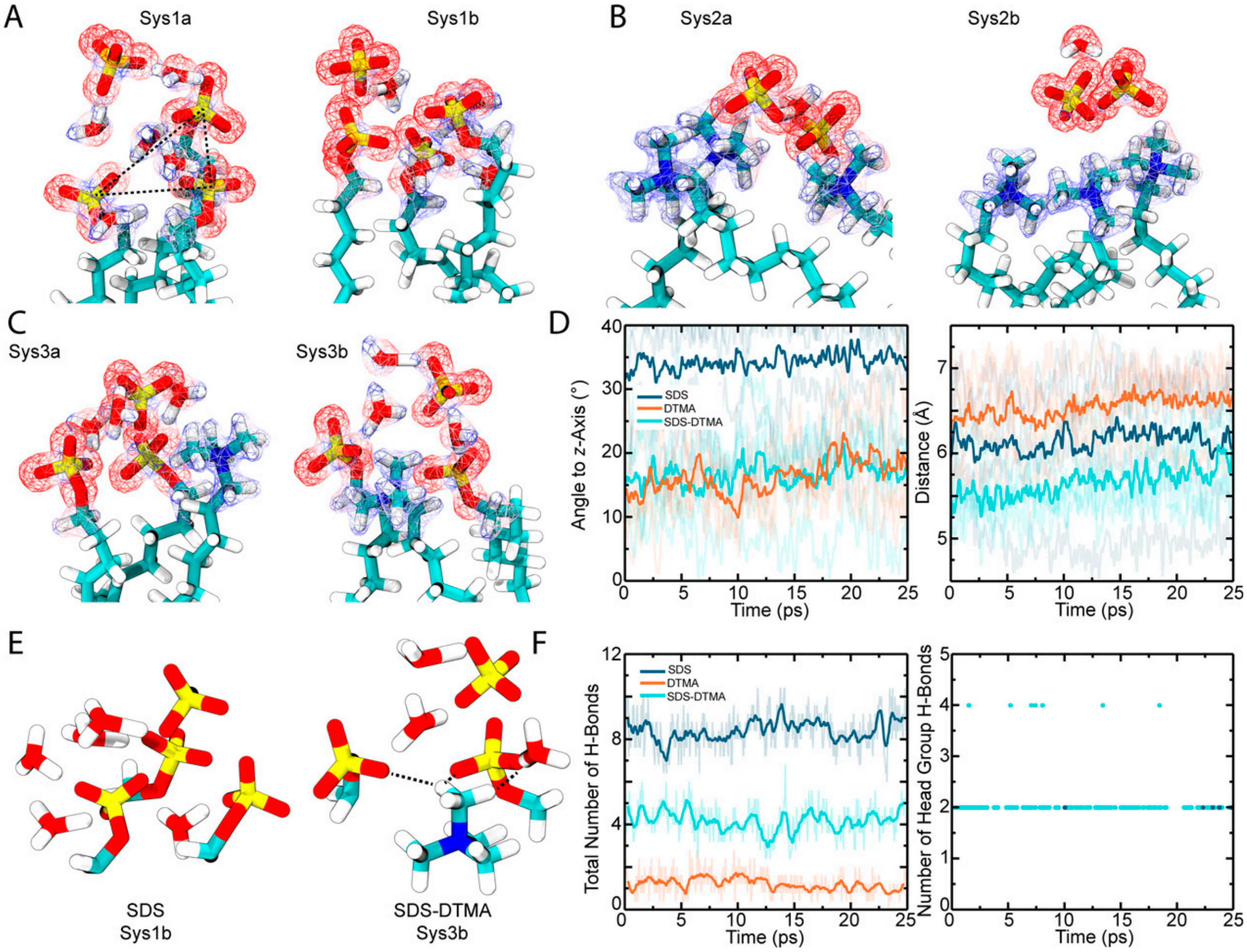

**FIG. 9.** Modeling and simulation of SDS and DTMA membranes. The final frame of the QM/MM simulations shows the electron density for the **(A)** SDS-only, **(B)** DTMA-only, and **(C)** SDS-DTMA systems. For clarity, only the lipids and solvent molecules associated with the QM region are shown. Dashed lines in **(A)** show the distances between head groups and the plane formed by the three head groups. **(D)** The angle between the axis perpendicular to the plane formed by the three head groups (left panel), and the average distances between head groups (right panel) are shown for all systems over the 5 25 ps subtrajectories. Individual subtrajectories are shown as transparent, while the average for each system is opaque. **(E)** Snapshot of the SDS-only system (left) and SDS-DTMA system (right), showing detail of the interaction between solvent molecules and head groups. Dashed lines indicate hydrogen bonds between head groups. **(F)** Total number of hydrogen bonds between all molecules showing the average of all five 25 ps subtrajectories (left panel; 0.6 ps running average shown in dark lines) and hydrogen bonds between head groups in all five subtrajectories (right panel). The results of these simulations provide a potential mechanistic explanation for the robustness of SDS-DTMA membranes in sulfuric acid. QM/MM, quantum mechanical/molecular mechanic.

## 4. Discussion

Stable membranes that can form vesicles are likely to be an essential requirement for cellular life, no matter its chemical makeup. Water is highly favored as a medium for life because, in addition to being an excellent solvent for polar molecules and salts, many molecules are insoluble in water, which allows for the hydrophobic effect that drives the formation and the stability of membranes (Pohorille and Pratt, 2012; Pratt and Pohorille, 1992; Tanford, 1978). However, the ability to facilitate the formation of membrane structures is not a unique property of water. For example, ammonia (Griffin *et al.*, 2015) and hydrogen fluoride (Roth *et al.*, 1995) also have the dense hydrogen bonding networks required to drive a solvophobic effect in the liquid state (reviewed in Bains *et al.*, 2024a). Here, we show the unexpected stability of complex membranous structures in another polar solvent: concentrated sulfuric acid.

Concentrated sulfuric acid as a planetary solvent could be widespread on exoplanets (Ballesteros *et al.*, 2019), either on exo-Venuses or on other rocky planets that are desiccated as a result of the stellar activity of their host star (Ostberg *et al.*, 2023). Concentrated sulfuric acid is also present in our immediate planetary vicinity, as a dominant liquid in the clouds of Venus (Titov *et al.*, 2018), which further emphasizes its importance for planetary science, planetary habitability, and astrobiology.

Our results suggest that the membranous structures observed in concentrated sulfuric acid are apparently resistant to sulfuric acid degradation and in some cases are dynamic. We cannot rule out that membrane dynamics through time could be due in part to some chemical degradation of the lipids. We emphasize that even if in some cases the tested lipids are reactive in concentrated sulfuric acid (Supplementary Fig. S4), this reactivity does not prevent the formation of higher-order structures. It is also possible that membranes may protect otherwise reactive lipids from degradation in concentrated sulfuric acid. Further work will be required to explore this possibility. Some of the tested lipids, for example, decylphosphonate, exhibited larger aggregate sizes compared to other samples, which suggests that the properties of a membrane could be tuned in sulfuric acid by changing the membrane lipid composition. This observation has implications for the potential biological plausibility of such membranes, as it is analogous to the characteristics of biological membranes in water.

Confocal microscopy showed the formation of vesicles and other structures in solution. The microscopy also revealed that, once formed, the lipid membranes adopt a variety of morphologies and can undergo structural changes, such as elongation up to tens of microns in length, often with associated pearling along with the elongated segment. The pearling behavior is especially noteworthy as it could facilitate division of membranous structures. Our results show that such division could in principle happen in concentrated sulfuric acid and that this solvent is compatible with complex and dynamic membranous systems.

Our experimental results are also in agreement with the QM/MM simulations. The simulations showed that in the SDS-DTMA lipid membrane system, the formation of a robust network of charge–charge interactions between both the head groups themselves and the solvent efficiently prevent the sulfuric acid solvent from passing through the SDS-DTMA membrane. This result raises the possibility that the SDS-DTMA membrane (or other similar lipid systems) could be robust enough to retain an aqueous vesicle lumen, or at least a distinct internal solvent environment, while keeping sulfuric acid solvent molecules from penetrating the membrane. Such membrane characteristics could in principle allow for protection and retention of acid-labile components, such as DNA, RNA, or proteins, within the vesicle.

By demonstrating the stability of lipid membranes in this aggressive solvent, we have taken a step forward in exploring the potential habitability of the concentrated sulfuric acid cloud environment on Venus. Future work should consider the experimental verification of the QM/MM simulations, as well as laboratory testing of the leakage of lipid membranes resistant to concentrated sulfuric acid.

## Acknowledgments

The authors are grateful to Andy Yu Zhi Li for assistance with confocal imaging. Simulations were performed using HPC resources on the Midway3 supercomputer at the Research Computing Center at the University of Chicago. This article was prepared without the use of artificial intelligence software.

## Authors' Contributions

Conceptualization: D.D., C.N., W.B., J.J.P., and S.S.; methodology: D.D., C.N., and C.K.K.; software: C.N.; formal analysis: D.D. and C.N.; investigation: D.D. and C.N.; writing—original draft preparation: D.D., C.N., and J.J.P.; writing—review and editing: D.D., C.N., J.J.P., S.S., W.B., J.W.S., and C.K.K.; supervision: S.S., D.D., and J.J.P.; and funding acquisition: S.S., J.W.S., and D.D. All authors have read and agreed to the published version of the article.

## Data Availability

Raw data and all data analysis files used to generate the figures are available on the Zenodo depository at https://zenodo.org/records/13628732.

## Author Disclosure Statement

The authors declare no conflicts of interest.

## Funding Information

Partial funding for the presented work came from MIT and Breakthrough Initiatives. The writing and editing of the article were supported in part by a Marie Skłodowska-Curie FRIAS COFUND Fellowship (to D.D.) at the Freiburg Institute for Advanced Studies (European Union Horizon 2020 research and innovation program under the Marie Skłodowska-Curie grant agreement No. 754340). J.W.S. is an investigator of the Howard Hughes Medical Institute.

## Supplementary Material

Supplementary Data
Supplementary Figure S1
Supplementary Figure S2
Supplementary Figure S3
Supplementary Figure S4
Supplementary Figure S5
Supplementary Figure S6
Supplementary Figure S7

## References

Ahrer E-M, Alderson L, Batalha NM, et al.; JWST Transiting Exoplanet Community Early Release Science Team. Identification of carbon dioxide in an exoplanet atmosphere. *Nature* 2023;614(7949):649–652.

Albright LF, Houle L, Sumutka AM, et al. Alkylation of Isobutane with Butenes: Effect of sulfuric acid compositions. *Ind Eng Chem Proc Des Dev* 1972;11(3):446–450.

Apel CL, Deamer DW, Mautner MN. Self-Assembled vesicles of monocarboxylic acids and alcohols: Conditions for stability and for the encapsulation of biopolymers. *Biochim Biophys Acta* 2002;1559(1):1–9.

Bains W, Petkowski JJ, Rimmer PB, et al. Production of ammonia makes venusian clouds habitable and explains observed cloud-level chemical anomalies. *Proc Natl Acad Sci* 2021a;118(52).

Bains W, Petkowski JJ, Rimmer PB, et al. Production of ammonia makes venusian clouds habitable and explains observed cloud-level chemical anomalies. *Proc Natl Acad Sci* 2021b;118(52):e2110889118.

Bains W, Petkowski JJ, Seager S. Alternative solvents for life: Framework for evaluation, current status and future research. *Astrobiology* 2024a;24(12):1231–1256.

Bains W, Petkowski JJ, Seager S. Venus' Atmospheric chemistry and cloud characteristics are compatible with Venusian life. *Astrobiology* 2024b;24(4):371–385; doi: 10.1089/ast.2022.0113

Ballesteros FJ, Fernandez-Soto A, Martínez VJ. Diving into exoplanets: Are water seas the most common? *Astrobiology* 2019;19(5):642–654; doi: 10.1089/ast.2017.1720

Beyer KD, Ravishankara AR, Lovejoy ER. Measurements of UV refractive indices and densities of H2SO4/H2O and H2SO4/HNO3/H2O solutions. *J Geophys Res* 1996;101(D9): 14519–14524.

Caschera F, de la Serna JB, Loffler PMG, et al. Stable Vesicles composed of Monocarboxylic or Dicarboxylic fatty acids and Trimethylammonium Amphiphiles. *Langmuir* 2011;27(23): 14078–14090.

Christiansen JL. Five thousand exoplanets at the NASA exoplanet archive. *Nat Astron* 2022;6(5):516–519.

Deamer D. Membranes and the origin of life: A century of conjecture. *J Mol Evol* 2016;83(5–6):159–168.

Gebicki JM, Hicks M. Ufasomes are stable particles surrounded by unsaturated fatty acid membranes. *Nature* 1973;243(5404): 232–234.

Griffin JM, Atherton JH, Page MI. Micelle formation in liquid ammonia. *J Org Chem* 2015;80(14):7033–7039.

Grinspoon DH, Bullock MA. Astrobiology and venus exploration. *Geophys Monogr Geophys Union* 2007;176:191.

Habeeb A. The reaction of sulphuric acid with lysozyme and horse globin. *Can J Biochem Physiol* 1961;39(1):31–43.

Hallsworth JE, Koop T, Dallas TD, et al. Water Activity in Venus's uninhabitable clouds and other planetary atmospheres. *Nat Astron* 2021;5(7):665–675.

Hargreaves WR, Deamer DW. Liposomes from ionic, single-chain Amphiphiles. *Biochemistry* 1978;17(18):3759–3768.

Huang J, Mackerell AD. CHARMM36 All-Atom additive protein force field: Validation based on comparison to NMR data. *J Comput Chem* 2013;34(25):2135–2145; doi: 10.1002/jcc.23354

Humphrey W, Dalke A, Schulten K. VMD: Visual Molecular dynamics. *J Mol Graph* 1996;14(1):33–38; doi: 10.1016/0263-7855(96)00018-5

Jo S, Kim T, Iyer VG, et al. CHARMM-GUI: A web-based graphical user interface for CHARMM. *J Comput Chem* 2008;29(11):1859–1865; doi: 10.1002/JCC.20945

Kanicky JR, Shah DO. Effect of premicellar aggregation on the p K a of fatty acid soap solutions. *Langmuir* 2003;19(6): 2034–2038.

Kotsyurbenko OR, Cordova JA, Belov AA, et al. Exobiology of the Venusian clouds: New insights into habitability through terrestrial models and methods of detection. *Astrobiology* 2021;21(10):1186–1205.

Kotsyurbenko OR, Kompanichenko VN, Brouchkov AV, et al. Different Scenarios for the origin and the subsequent succession of a hypothetical microbial community in the cloud layer of venus. *Astrobiology* 2024;24(4):423–441.

Liler M. *Reaction Mechanisms in Sulphuric Acid and Other Strong Acid Solutions*. Academic Press: London; 1971.

Limaye SS, Mogul R, Smith DJ, et al. Venus' Spectral signatures and the potential for life in the clouds. *Astrobiology* 2018;18(9):1181–1198.

Liu B, Gao M, Li H, et al. Model of protocell compartments–Dodecyl Hydrogen Sulfate vesicles. *Phys Chem Chem Phys* 2018;20(3):1332–1336.

Mansy SS. Model protocells from single-chain lipids. *Int J Mol Sci* 2009;10(3):835–843.

McCulloch L. Black "Soap" Films of fatty acids in sulfuric acid. *J Am Chem Soc* 1946;68(12):2735–2736.

Melo MCR, Bernardi RC, Rudack T, et al. NAMD Goes Quantum: An integrative suite for QM/MM simulations. *Nat Methods* 2018;15(5):351–354; doi: 10.1038/NMETH.4638

Menger FM, Jerkunica JM. Aggregation in strong acid. A micelle of carbonium ions. *J Am Chem Soc* 1979;101(7):1896–1898.

Miron S, Lee RJ. Molecular structure of conjunct polymers. *J Chem Eng Data* 1963;8(1):150–160.

Mogul R, Limaye SS, Lee YJ, et al. Potential for Phototrophy in Venus' clouds. *Astrobiology* 2021a;21(10):1237–1249; doi: 10.1089/ast.2021.0032

Mogul R, Limaye SS, Way MJ, et al. Venus' Mass Spectra Show Signs of Disequilibria in the Middle Clouds. *Geophys Res Lett* 2021b;48(7):e2020GL091327.

Morowitz H, Sagan C. Life in the clouds of Venus? *Nature* 1967;215(5107):1259–1260; doi: 10.1038/2151259a0

Müller A. The Behaviour of Surfactants in concentrated acids Part II. Formation of micelles from cationic surfactants in concentrated sulphuric acid. *Colloids and Surfaces* 1991a; 57(2):219–226.

Müller A. The Behaviour of surfactants in concentrated acids part III. Surface activity of Cationics in concentrated sulphuric acid. *Colloids and Surfaces* 1991b;57(2):227–238.

Müller A. The Behaviour of surfactants in concentrated acids part IV. Investigation of the synergism of binary cationic/cationic mixtures in concentrated sulphuric acid. *Colloids and Surfaces* 1991c;57(2):239–247.

Müller A and Burchard W. Structure formation of surfactants in concentrated sulphuric acid: A light scattering study. *Colloid Polym Sci* 1995;273(9):866–875.

Müller A, Giersberg S. The Behaviour of surfactants in concentrated acids 5. Micellization and Interfacial activity of n-Alkyltrimethylammonium Bromides in methane sulphonic acid and concentrated sulphuric acid. *Colloids and Surfaces* 1991;60: 309–324.

Müller A, Giersberg S. The behaviour of surfactants in concentrated acids 6. Micellization and interfacial behaviour of n-Hexadecyltrimethylammonium bromide in mixtures of methane sulphonic acid and concentrated sulphuric acid. *Colloids and Surfaces* 1992;69(1):5–14.

Müller A, Miethchen R. Amphiphile Aggregation in Konzentrierten Säuren Grenzflächenaktivität von N-(N-Alkyl)-pyridinium-Salzen in Schwefelsäure. *J Prakt Chem* 1988;330(6): 993–1005.

Neese F, Wennmohs F, Becker U, et al. The ORCA Quantum chemistry program package. *J Chem Phys* 2020;152(22): 224108; doi: 10.1063/5.0004608

Ostberg C, Kane SR, Li Z, et al. The demographics of terrestrial planets in the Venus zone. *AJ* 2023;165(4):168.

Patel MR, Mason JP, Nordheim TA, et al. Constraints on a potential aerial biosphere on Venus: II. Ultraviolet radiation. *Icarus* 2022;373:114796; doi: 10.1016/j.icarus.2021.114796

Petkowski JJ, Seager MD, Bains W, et al. General instability of dipeptides in concentrated sulfuric acid as relevant for the venus cloud habitability. *Sci Rep* 2024a;14(1):17083.

Petkowski JJ, Seager MD, Bains W, et al. Mechanism for peptide bond Solvolysis in 98% w/w concentrated sulfuric acid. *ACS Omega* 2025;10(9):9623–9629; doi: 10.1021/acsomega.4c10873in press;

Petkowski JJ, Seager S, Grinspoon DH, et al. Astrobiological potential of Venus atmosphere chemical anomalies and other unexplained cloud properties. *Astrobiology* 2024b;24(4):343–370.

Phillips JC, Hardy DJ, Maia JDC, et al. Scalable molecular dynamics on CPU and GPU architectures with NAMD. *J Chem Phys* 2020;153(4):44130; doi: 10.1063/5.0014475

Pohorille A, Pratt LR. Is water the universal solvent for life? *Orig Life Evol Biosph* 2012;42(5):405–409.

Pratt LR, Pohorille A. Theory of hydrophobicity: Transient cavities in molecular liquids. *Proc Natl Acad Sci U S A* 1992; 89(7):2995–2999.

Reitz HC, Ferrel RE, Fraenkel-Conrat H, et al. Action of sulfating agents on proteins and model substances. I. Concentrated sulfuric acid. *J Am Chem Soc* 1946;68(6):1024–1031.

Rhodes FH, Barbour CB. The viscosities of mixtures of sulfuric acid and water. *Ind Eng Chem* 1923;15(8):850–852.

Roth U, Paulus O, Menyes U. Surface activity of Amphiphiles in hydrogen fluoride—water solutions. *Colloid Polym Sci* 1995;273(8):800–806.

Sagnella DE, Voth GA. Structure and dynamics of hydronium in the ion channel gramicidin A. *Biophys J* 1996;70(5): 2043–2051; doi: 10.1016/S0006-3495(96)79773-4

Schulze-Makuch D, Irwin LN. The prospect of alien life in exotic forms on other worlds. *Naturwissenschaften* 2006; 93(4):155–172.

Schumacher M, Günther H. Beiträge Zur 15N-NMR-Spektroskopie Protonierung Und Tautomerie in Purinen: Purin Und 7-und 9-Methylpurin. *Chem Ber* 1983;116(5):2001–2014.

Seager S, Petkowski JJ, Gao P, et al. The Venusian lower atmosphere haze as a depot for desiccated microbial life: A proposed life cycle for persistence of the Venusian aerial biosphere. *Astrobiology* 2021;21(10):1206–1223.

Seager S, Petkowski JJ, Seager MD, et al. Stability of nucleic acid bases in concentrated sulfuric acid: Implications for the habitability of Venus' clouds. *Proc Natl Acad Sci U S A* 2023;120(25):e2220007120; doi: 10.1073/pnas.2220007120

Seager S, Petkowski JJ, Seager MD, et al. Year-Long stability of nucleic acid bases in concentrated sulfuric acid: Implications for the persistence of organic chemistry in Venus' clouds. *Life (Basel)* 2024b;14(5):538; doi: 10.3390/life14050538

Seager MD, Seager S, Bains W, et al. Stability of 20 Biogenic amino acids in concentrated sulfuric acid: Implications for the habitability of Venus' clouds. *Astrobiology* 2024a;24(4):386–396.

Spacek J, Rimmer P, Owens GE, et al. Production and reactions of organic molecules in clouds of Venus. *ACS Earth Space Chem* 2024;8(1):89–98; doi: 10.1021/acsearthspacechem.3c00261

Steigman J, Shane N. Micelle formation in concentrated sulfuric acid as solvent. *J Phys Chem* 1965;69(3):968–973.

Sure R, Grimme S. Corrected small basis set Hartree-Fock method for large systems. *J Comput Chem* 2013;34(19): 1672–1685; doi: 10.1002/JCC.23317

Tanford C. The hydrophobic effect and the organization of living matter. *Science* 1978;200(4345):1012–1018.

Titov DV, Ignatiev NI, McGouldrick K, et al. Clouds and Hazes of Venus. *Space Sci Rev* 2018;214(8):1–61.

Torn RD, Nathanson GM. Surface tensions and surface segregation of N-Butanol in sulfuric acid. *J Phys Chem B* 2002; 106(33):8064–8069.

Tsai S-M, Lee EKH, Powell D, et al. Photochemically produced SO2 in the atmosphere of WASP-39b. *Nature* 2023;617 (7961):483–487.

Vanommeslaeghe K, MacKerell AD. Automation of the CHARMM General Force Field (CGenFF) I: Bond perception and atom typing. *J Chem Inf Model* 2012;52(12): 3144–3154; doi: 10.1021/CI300363C

Wagner R, von Philipsborn W. Protonierung von Purin, Adenin Und Guanin NMR.-Spektren Und Strukturen Der Mono-, Di-und Tri-Kationen. *Helv Chim Acta* 1971;54(6): 1543–1558.

Walde P, Wessicken M, Rädler U, et al. Preparation and characterization of vesicles from Mono-n-Alkyl Phosphates and Phosphonates. *J Phys Chem B* 1997;101(38):7390–7397.

Willcott MR. MestRe Nova. *J Am Chem Soc* 2009;131(36): 13180–13180; doi: 10.1021/ja906709t

Zhu TF, Szostak JW. Coupled growth and division of model protocell membranes. *J Am Chem Soc* 2009;131(15):5705–5713; doi: 10.1021/ja900919c

Address correspondence to:
*Daniel Duzdevich*
*Department of Chemistry*
*Searle Chemistry Laboratory*
*The University of Chicago, 5735 South Ellis*
*Avenue Chicago, IL 60637*
*USA*

*E-mail:* dduzdevich@uchicago.edu

**Abbreviations Used**

| | |
|---|---|
| BODIPY | = 4,4-difluoro-4-bora-3a,4a-diaza-s-indacene fluorescent dye |
| C10 and C18 | = 10-carbon and 18-carbon lipid tail length, respectively |
| DLS | = dynamic light scattering |
| DTMA | = decyltrimethylamine |
| fs | = femtosecond |
| HF-3c | = Hartree-Fock-3c |
| QM/MM | = quantum mechanical/molecular mechanical (simulation) |
| SDS | = (sodium) decylsulfate |
| SOS | = (sodium) octadecyl sulfate |
| TMO | = octadecyltrimethylammonium |
| v/v | = volume/volume |
| VMD | = visual molecular dynamics |

## Supplementary Information

# Simple lipids form stable higher-order structures in concentrated sulfuric acid

Daniel Duzdevich[1,2,†*], Collin Nisler[1,†], Janusz J. Petkowski[3,4], William Bains[5], Caroline K. Kaminsky[1], Jack W. Szostak[1,6], Sara Seager[7,8,9]

[1] Department of Chemistry, The University of Chicago, Searle Chemistry Laboratory, Chicago, IL 60637, USA
[2] Freiburg Institute for Advanced Studies, Albert-Ludwigs-Universität Freiburg, 79104 Freiburg im Breisgau, Germany
[3] Faculty of Environmental Engineering, Wroclaw University of Science and Technology, 50-370 Wroclaw, Poland
[4] JJ Scientific, Mazowieckie, Warsaw, Poland
[5] School of Physics and Astronomy, Cardiff University, 4 The Parade, Cardiff CF24 3AA, UK
[6] Howard Hughes Medical Institute, The University of Chicago, Chicago, IL 60637, United States of America
[7] Department of Earth, Atmospheric and Planetary Sciences, Massachusetts Institute of Technology, 77 Massachusetts Avenue, Cambridge, MA 02139, USA
[8] Department of Physics, Massachusetts Institute of Technology, 77 Massachusetts Avenue, Cambridge, MA 02139, USA
[9] Department of Aeronautics and Astronautics, Massachusetts Institute of Technology, 77 Massachusetts Avenue, Cambridge, MA 02139, USA

† Equal contribution
* To whom correspondence should be addressed. Email: dduzdevich@uchicago.edu

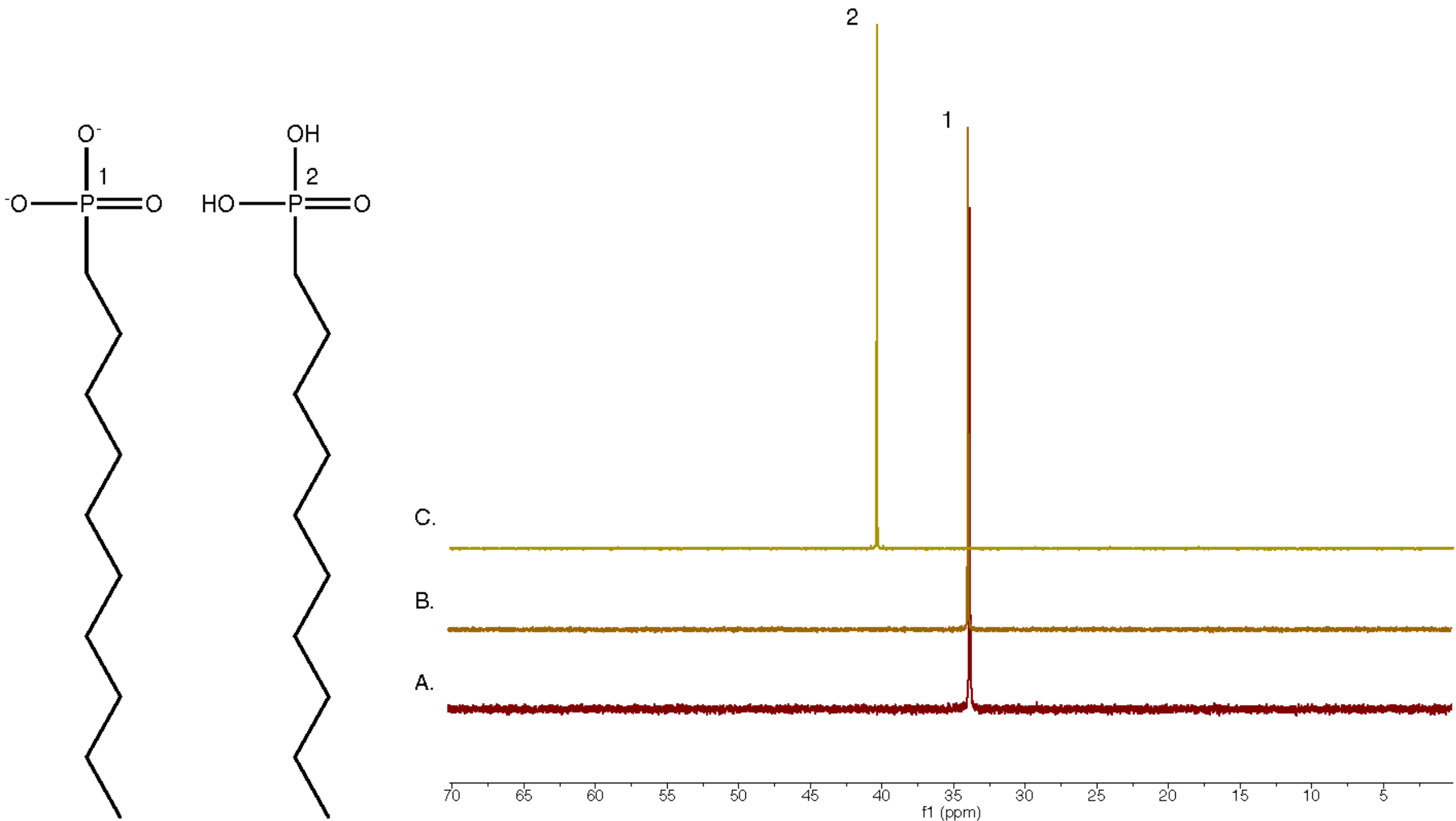

**Supplemental Figure 1. Decylphosphonate is resistant to 70% (v/v) sulfuric acid.** $^{31}P$ NMR spectra of 50 mM decylphosphonate incubated in (A) $D_2O$, (B) $D_2O$ + $H_2SO_4$, and (C) 30% $D_2O$ + 70% $H_2SO_4$ for one hour and extracted into $CDCl_3$. Note the absence of new peaks.

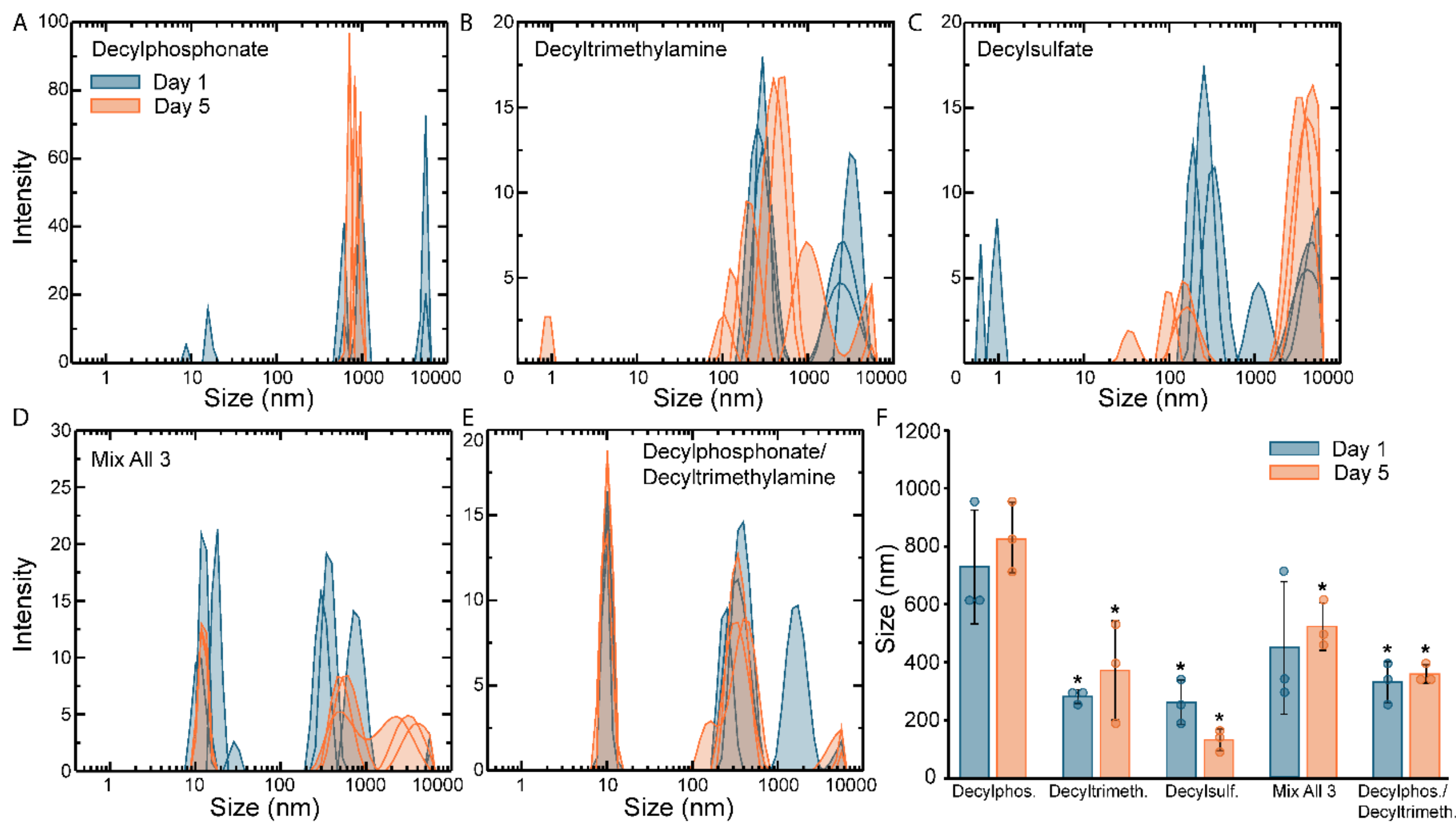

**Supplemental Figure 2. Initial DLS experiments for all three lipids.** (A-E) DLS readings of the indicated 25 mM lipids prepared in 70% sulfuric acid. All three measurements are shown for Day 1 (blue) and Day 5 (orange). (F) The average size of lipid structures for each lipid preparation is indicated by solid bars, individual values are indicated by circles, and standard deviations are in black. Sizes at peak intensity were taken for signals between 100 and 1000 nm. Values that are significantly different from decylphosonate are marked by asterisks, which were the only values that exhibited significant differences between all samples. These experiments indicate that all three lipids can form structures in concentrated sulfuric acid.

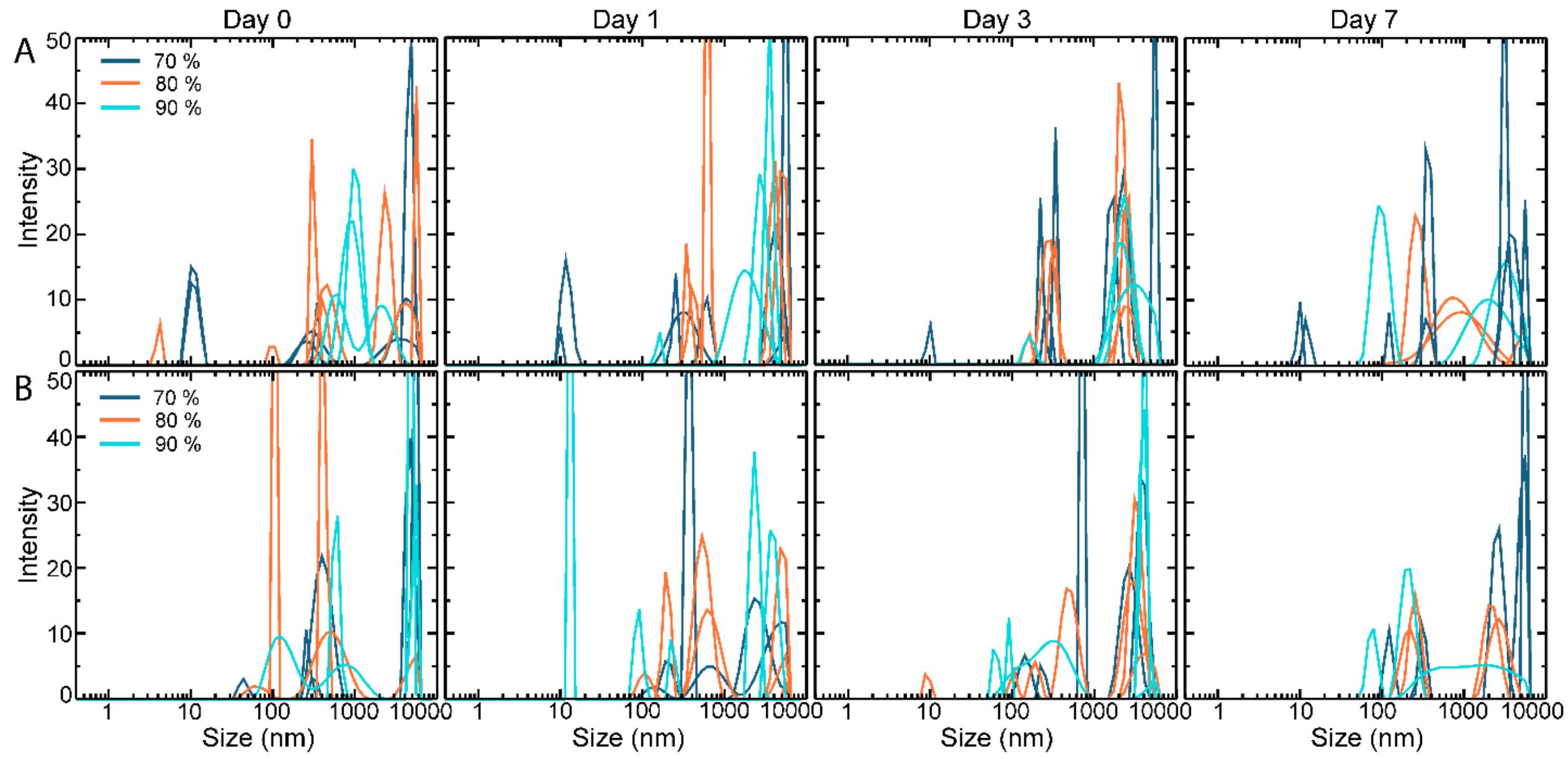

**Supplemental Figure 3. All DLS experiments for all DTMA-SDS and SOS-TMO lipids.** (A) DLS readings for DTMA-SDS in 70%, 80%, or 90% sulfuric acid at Day 0, Day 1, Day 3, and Day 7 shown in triplicate. (B) DLS readings for SOS-TMO in 70%, 80%, and 90% sulfuric acid at Day 0, Day 1, Day 3, and Day 7 shown in triplicate. While there is variability observed between and among samples, such variability is supported by the microscopy results that also demonstrates a heterogeneous mix of lipid structures.

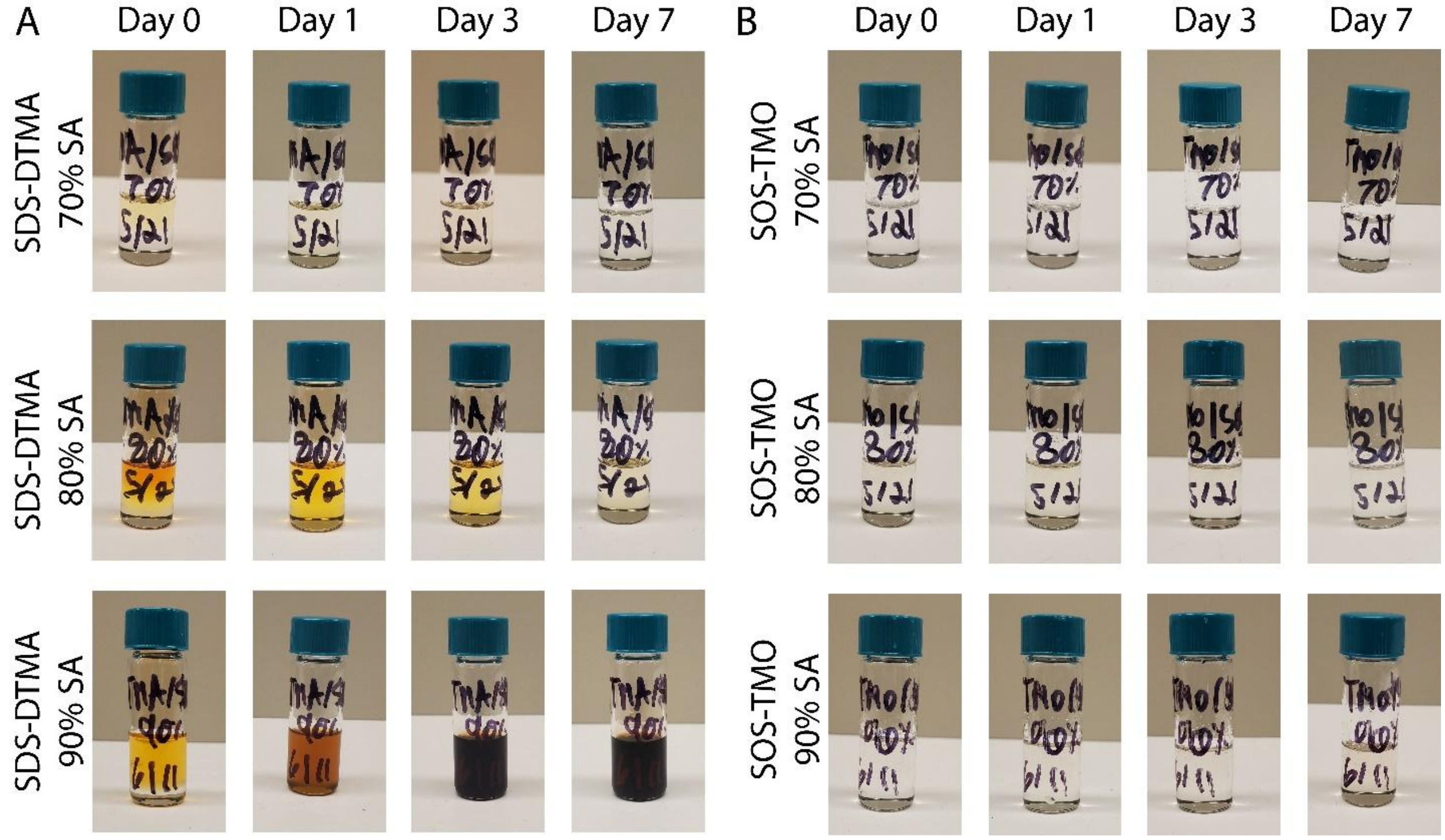

**Supplemental Figure 4. Visualization of sample turbidity over time.** Qualitative observation of the reaction between lipids and sulfuric acid for (A) 75 mM SDS-DTMA and (B) 1 mM SOS-TMO over the course of seven days. SDS-DTMA in 70% or 80% sulfuric acid becomes slightly clearer with time and exhibits a distinct chemistry at the surface. (This may in part be due to the bromide counterion ($Br^-$) of the DTMA salt forming bromine ($Br_2$) in the presence of concentrated sulfuric acid.) SDS-DTMA in 90% sulfuric acid becomes discolored after one day. In contrast, SOS-TMO does not change color with time, except in 90% sulfuric acid after 7 days, when it is slightly discolored. The concentration of SDS-DTMA lipids tested here is much higher than the concentration of SOS-TMO lipids (75 mM versus 1 mM). Therefore, much more material is available to react with the sulfuric acid in the case of SDS-DTMA. It is unclear what causes the discoloration. Note that the apparent reactivity of lipids in concentrated sulfuric acid does not inhibit the formation of the lipid structures (see Figure 5).

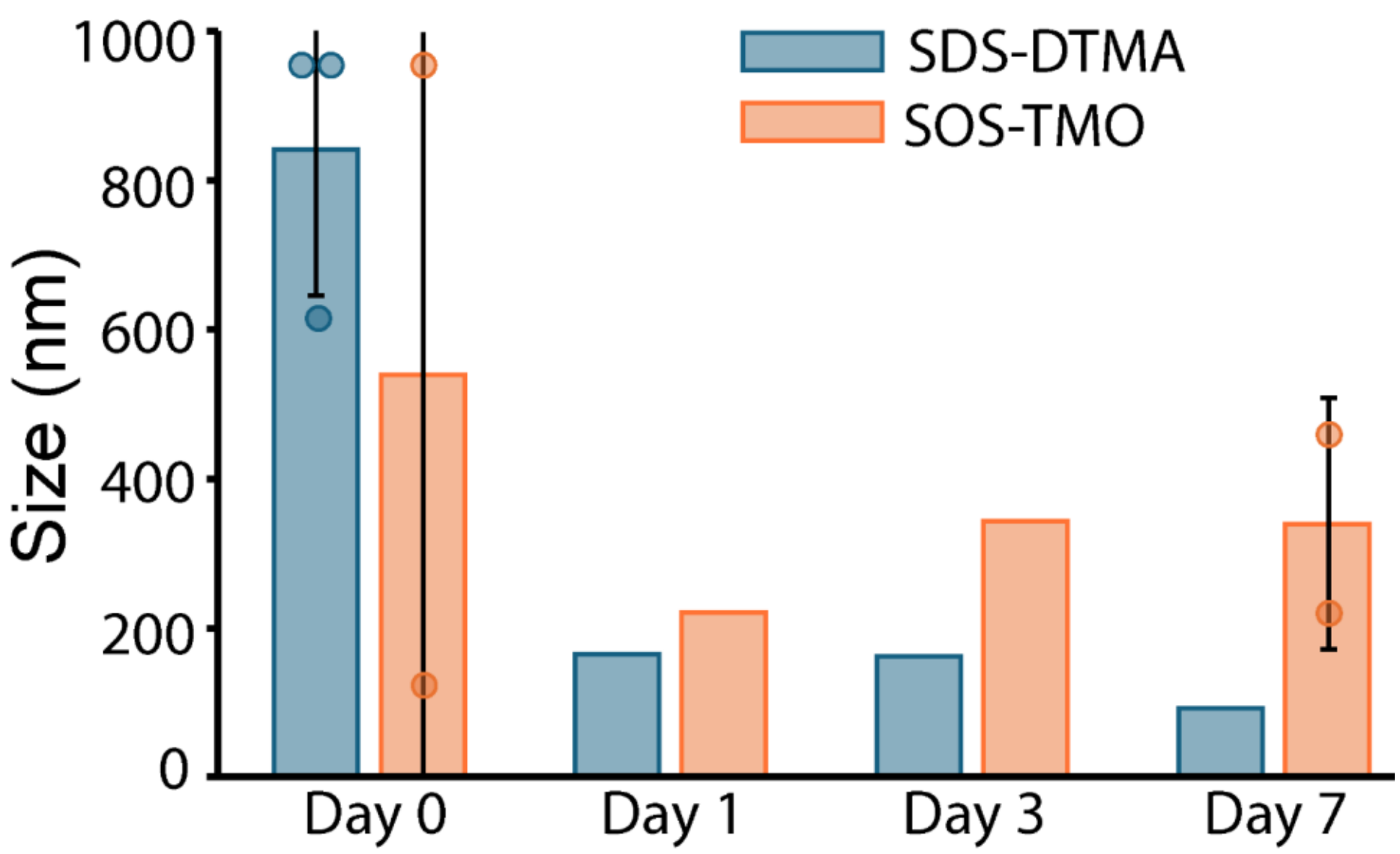

**Supplemental Figure 5. Lipids can form structures in 90% sulfuric acid.** DLS readings of 75 mM SDS-DTMA (blue) and 1 mM SOS-TMO (orange) lipids prepared in 90% sulfuric acid are shown hours after preparation (Day 0), at Day 1, Day 3, and Day 7. While signals were observed between 100 and 1000 nm in only one replicate at Day 1, Day 3, and SDS-DTMA Day 7, shown here, the signals observed indicate that lipid-based structures are possible in 90% sulfuric acid.

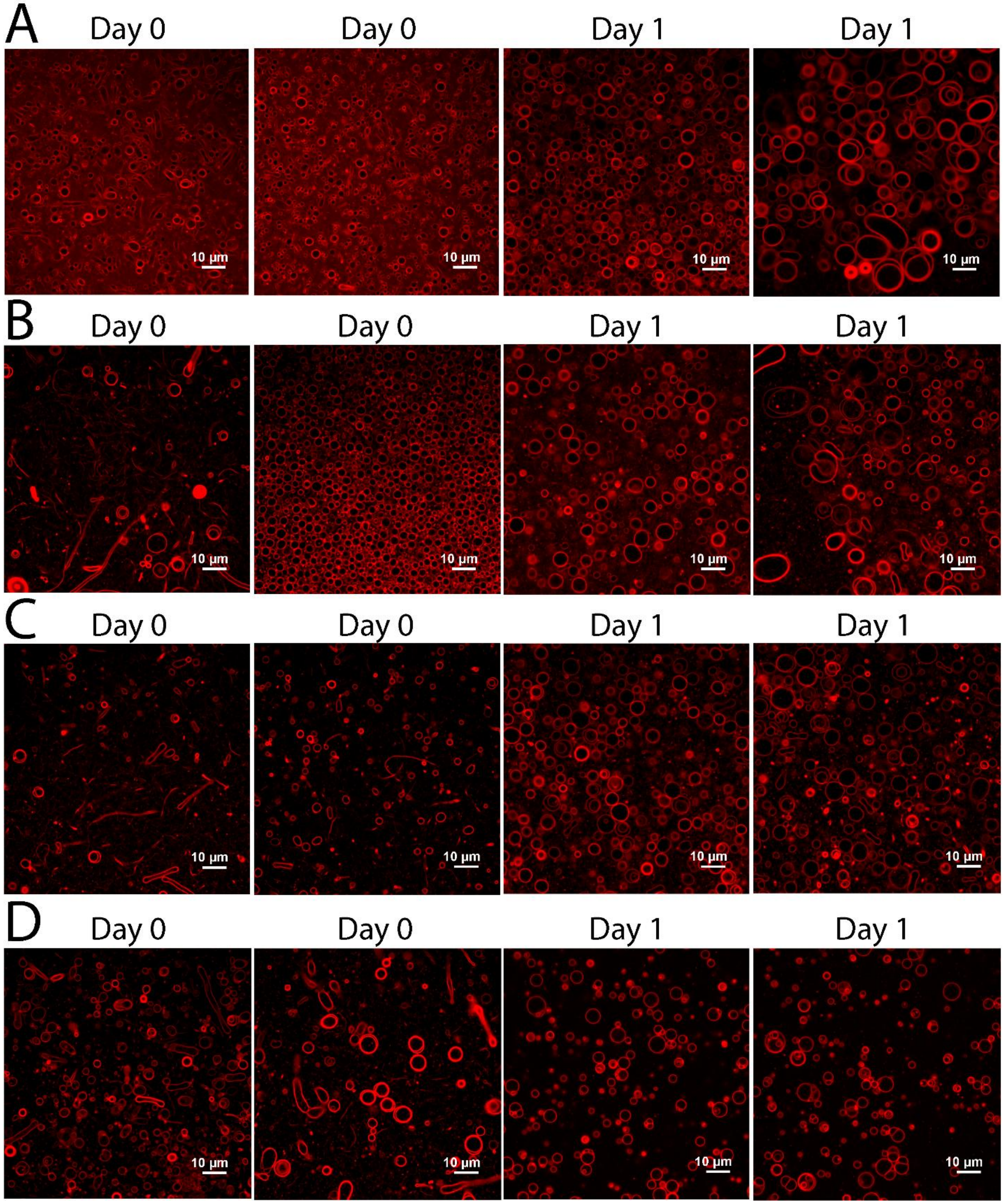

**Supplemental Figure 6. SDS-DTMA vesicles are unaffected by heat shock and osmotic shock.** Confocal images of BODIPY-stained SDS-DTMA lipids at Day 0 and Day 1 (A) after a 65˚C heat shock, or after addition of water where the total final sample volume is (B) 70%, (C) 80%, or (D) 90% newly added water. The lack of morphology changes in A-D indicates that changes observed upon the addition of sulfuric acid are likely due to the change in solvent properties and rather than released heat or osmotic shock.

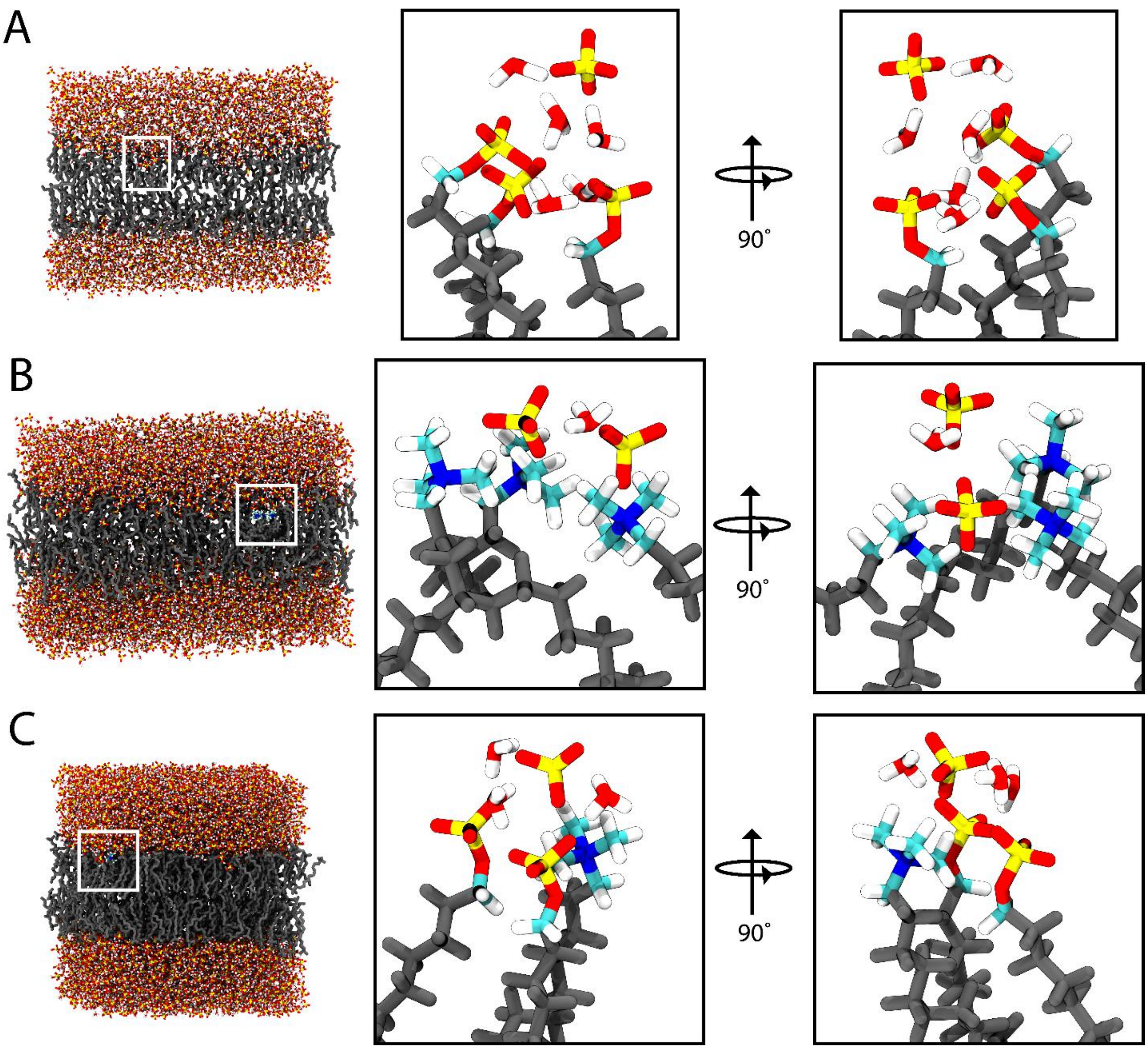

**Supplemental Figure 7. Overview of simulation systems.** Simulation systems (A) Sys1a, (B) Sys2a, and (C) Sys3a. Left panels show lipids in grey and all solvent molecules explicitly, cut along the *z*-axis (perpendicular to membrane surface) at the point of the QM region. Center panels show the QM region for each system in color with the MM part of the lipid in grey. Right panels show the view in the center panels rotated by 90˚.

## Appendix: Full-spectrum NMRs

(Figure 1)

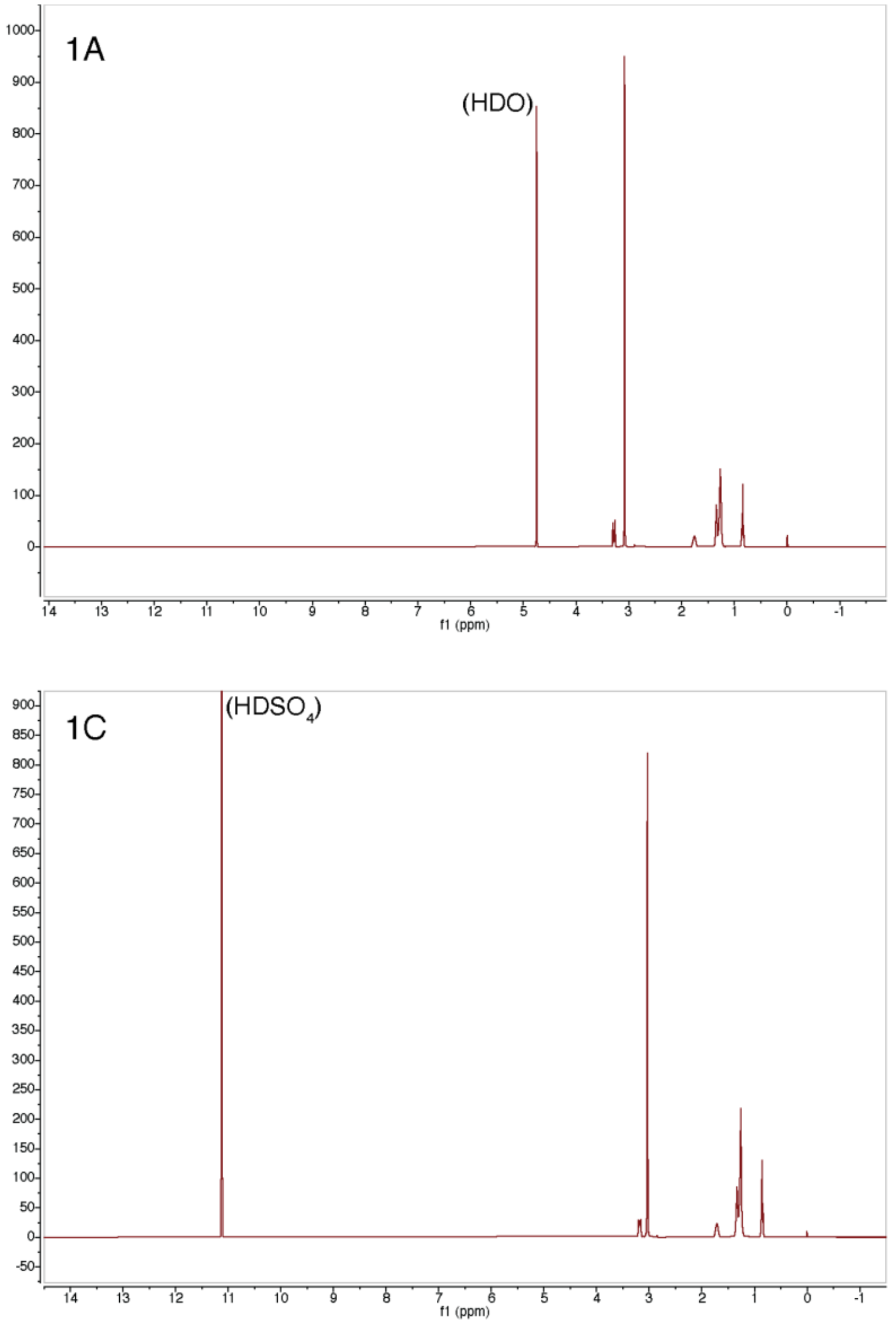

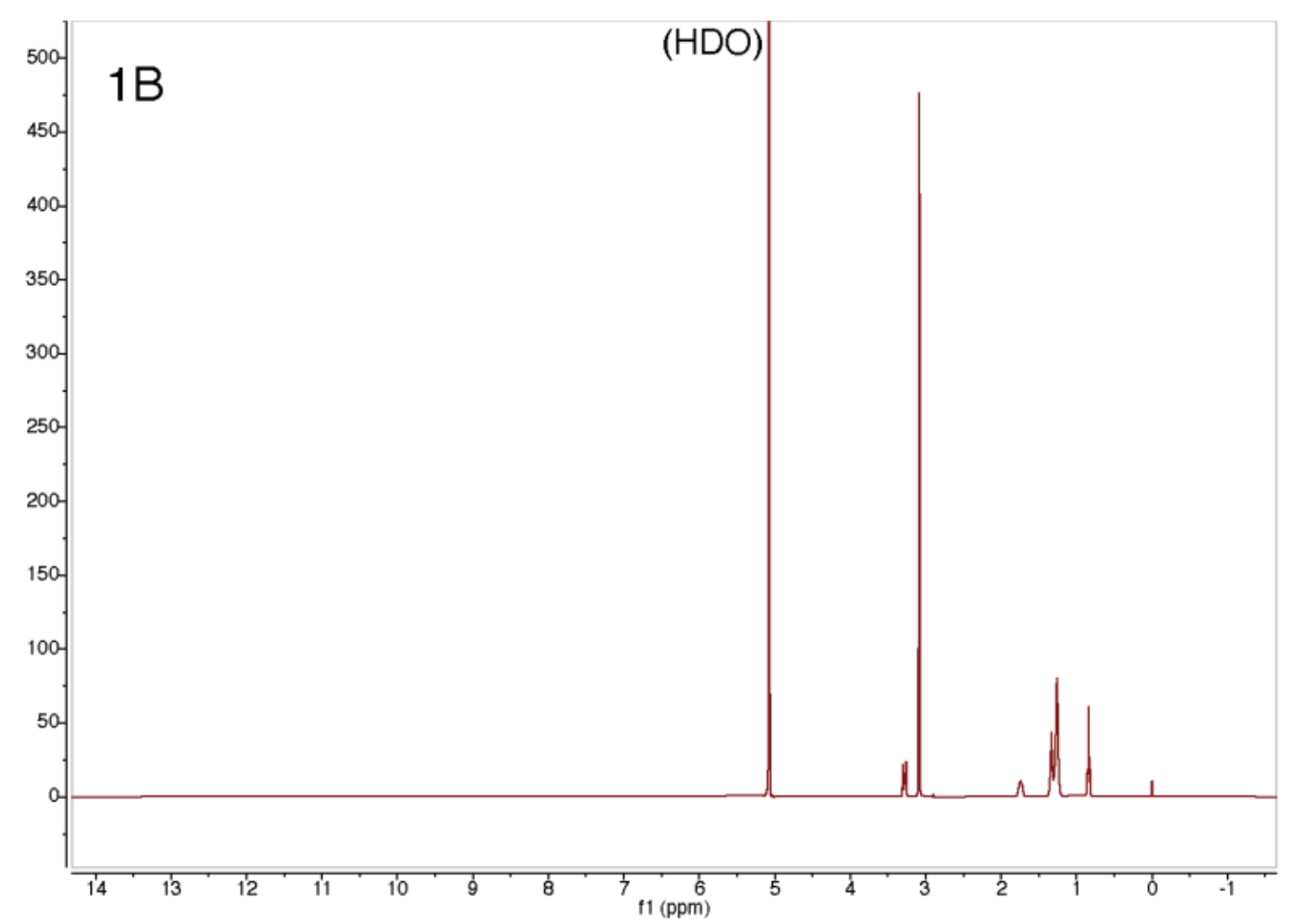

(Figures 2 & 3)

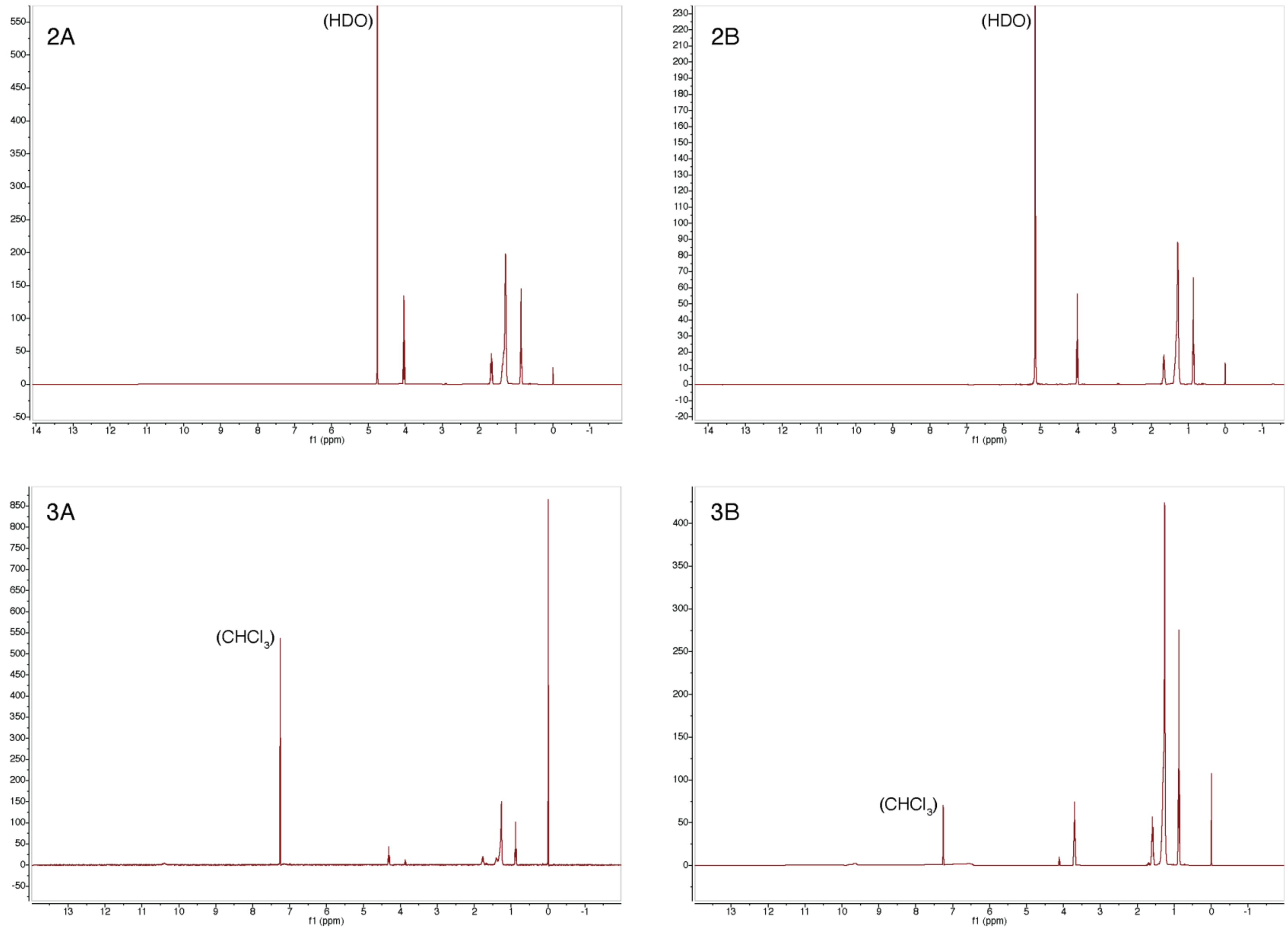

(Figure 4)

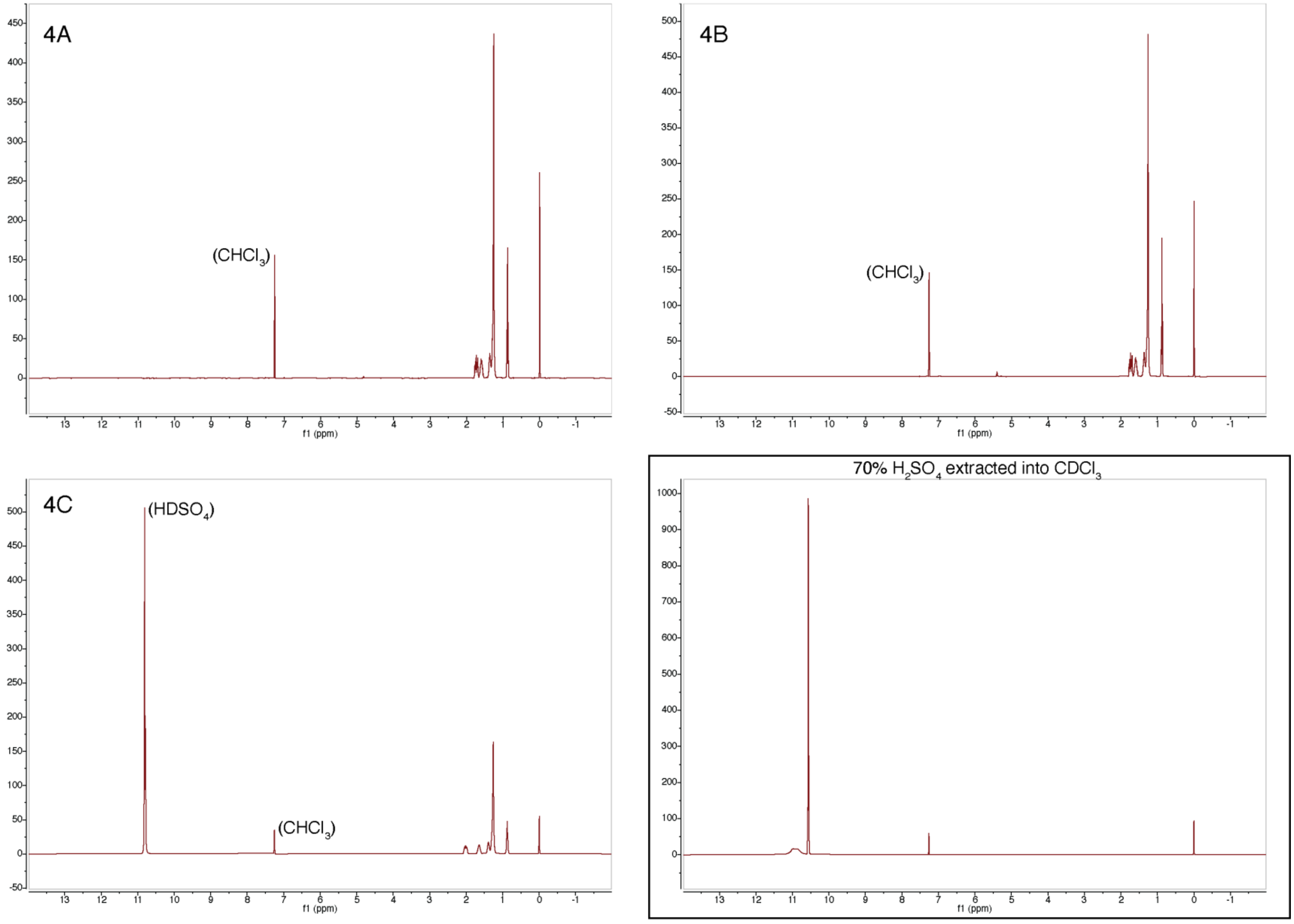

(Figure S1)

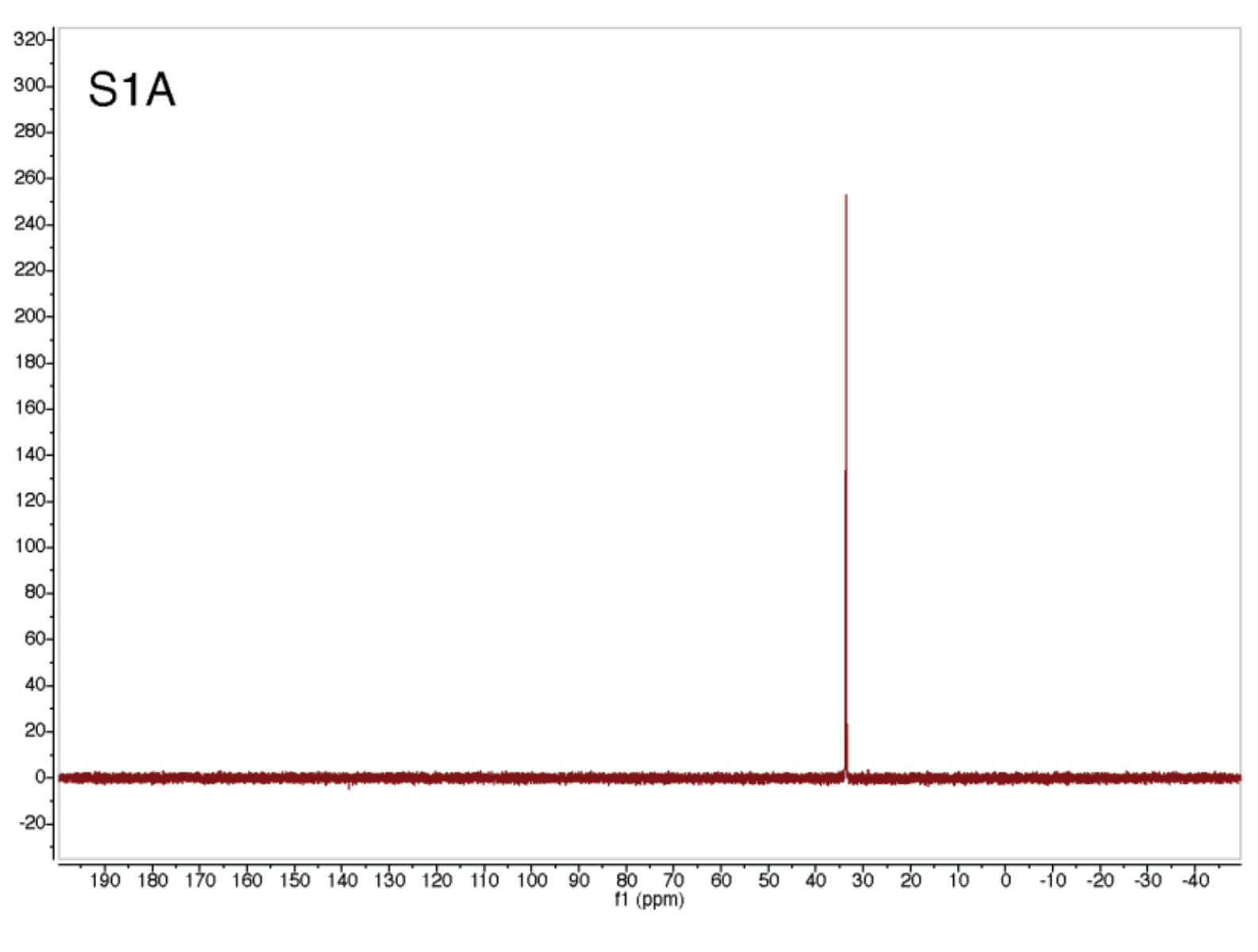

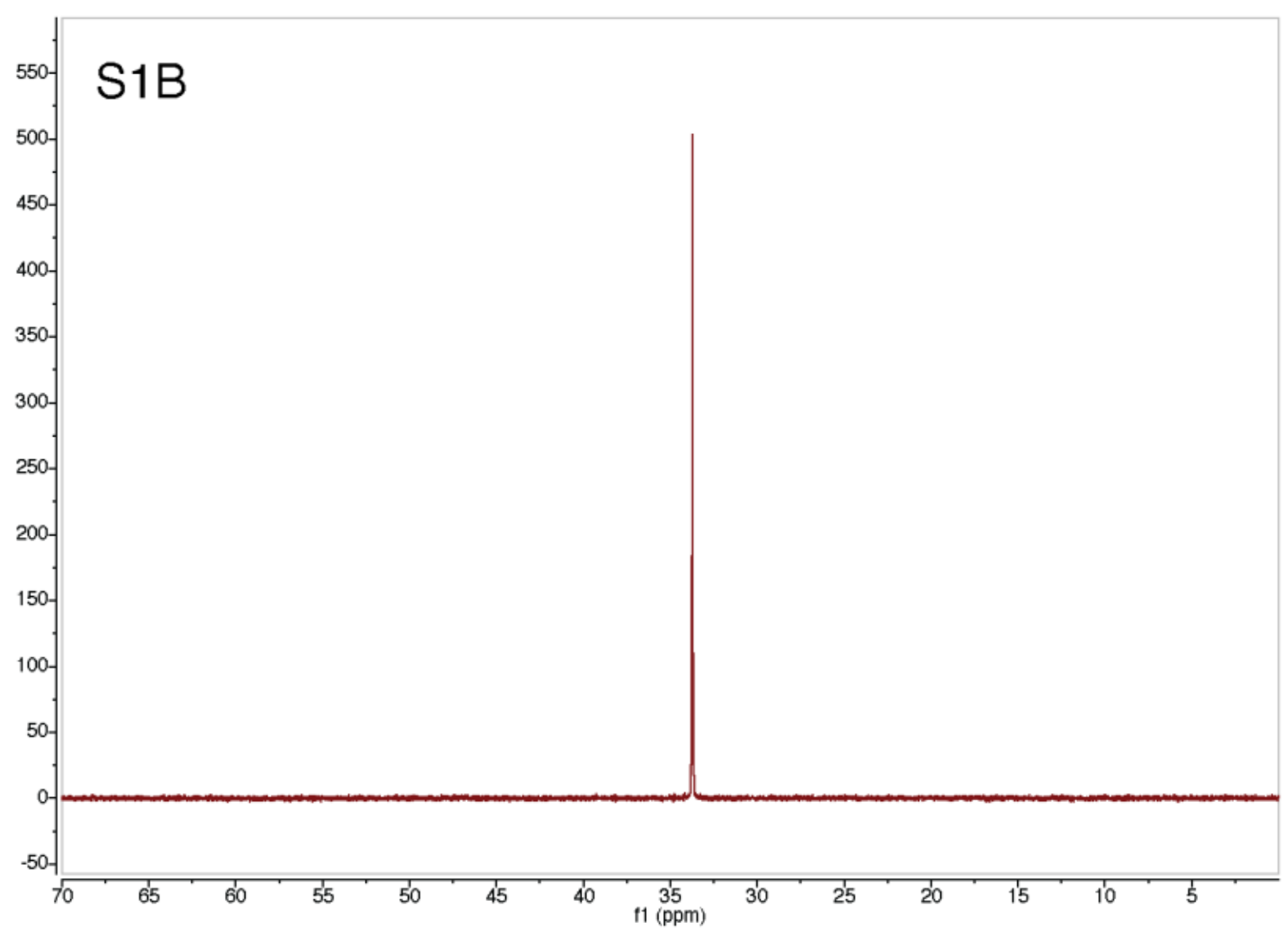

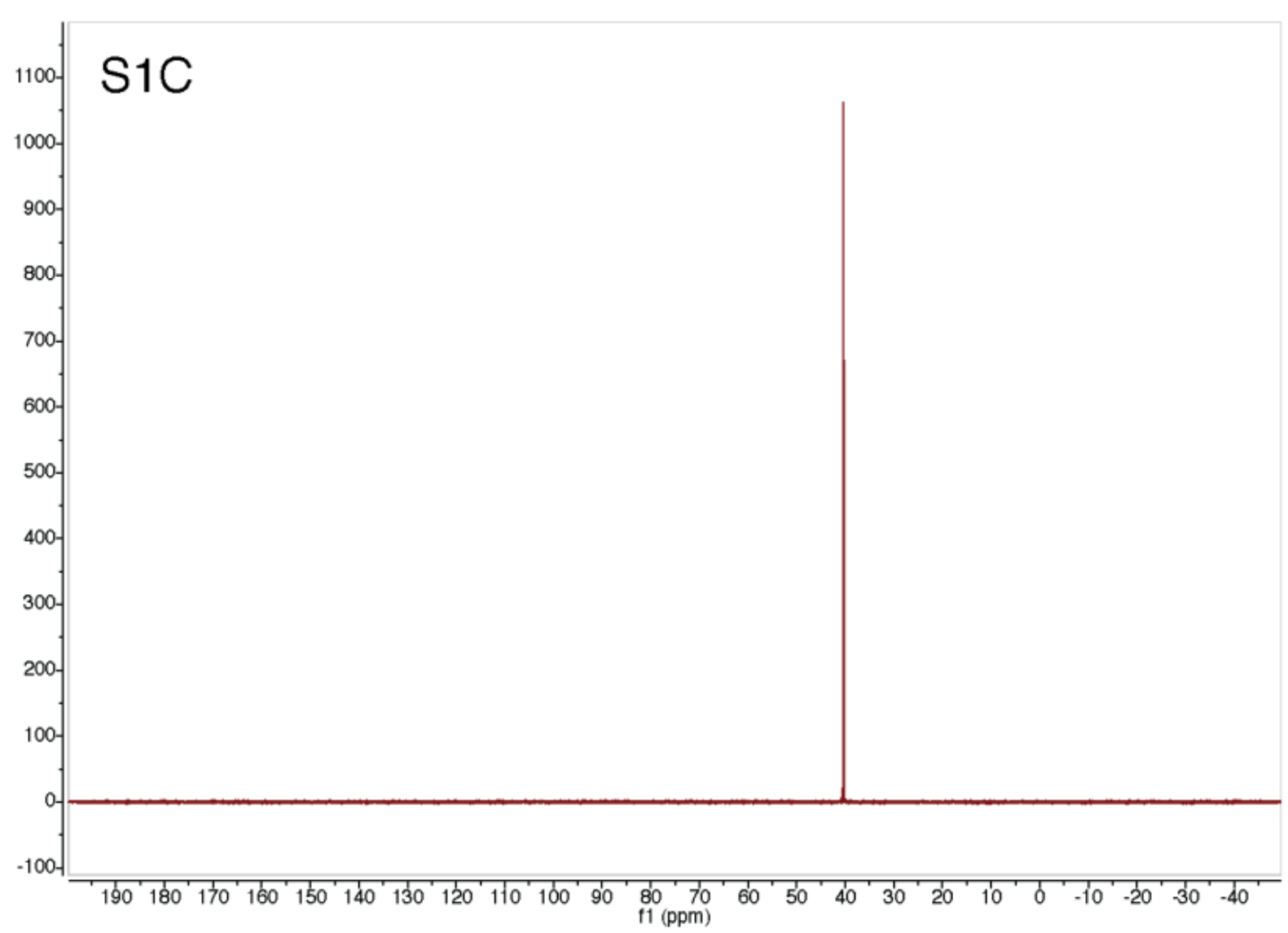